\documentclass[revtex4, 12pt,preprint,useAMS,usenatbib]{emulateapj}
\usepackage{graphicx}
\usepackage{setspace}
\usepackage{natbib}
\usepackage{color}
\usepackage{amsmath,amssymb}

\slugcomment{ApJ, accepted}
\shorttitle{Touching The Void: A Striking Drop in Stellar Halo Density Beyond 50 kpc}
\shortauthors{Deason, Belokurov, Koposov \& Rockosi}

\begin{document}

\title{Touching The Void: A Striking Drop in Stellar Halo Density Beyond 50 kpc}
\author{A. J. Deason\altaffilmark{1,4}, V. Belokurov\altaffilmark{2},  S. E. Koposov\altaffilmark{2,3}, C. M. Rockosi\altaffilmark{1}}

\altaffiltext{1}{Department of Astronomy and Astrophysics, University of California Santa Cruz, Santa Cruz, CA 95064, USA; alis@ucolick.org}
\altaffiltext{2}{Institute of Astronomy, Madingley Rd, Cambridge, CB3 0HA, UK}
\altaffiltext{3}{Moscow MV Lomonosov State University, Sternberg Astronomical Institute, Moscow 119992, Russia}
\altaffiltext{4}{Hubble Fellow}


\date{\today}

\begin{abstract}
We use A-type stars selected from Sloan Digital Sky Survey data release 9 photometry to measure the outer slope of the Milky Way stellar halo density profile beyond $50$ kpc. A likelihood-based analysis is employed that models the $ugr$ photometry distribution of blue horizontal branch (BHB) and blue straggler (BS) stars. In the magnitude range, $18.5 < g < 20.5$, these stellar populations span a heliocentric distance range of: $10 \lesssim D_{\rm BS}/\mathrm{kpc} \lesssim 75$,  $40 \lesssim D_{\rm BHB}/\mathrm{kpc} \lesssim 100$. Contributions from contaminants, such as QSOs, and the effect of photometric uncertainties, are also included in our modeling procedure.  We find evidence for a very steep outer halo profile, with power-law index $\alpha \sim 6$ beyond Galactocentric radii $r=50$ kpc, and even steeper slopes favored ($\alpha \sim 6-10$) at larger radii. This result holds true when stars belonging to known overdensities, such as the Sagittarius stream, are included or excluded. We show that, by comparison to numerical simulations, stellar halos with shallower slopes at large distances tend to have more recent accretion activity. Thus, it is likely that the Milky Way has undergone a relatively quiet accretion history over the past several Gyr. Our measurement of the outer stellar halo profile may have important implications for dynamical mass models of the Milky Way, where the tracer density profile is strongly degenerate with total mass-estimates.

\end{abstract}

\section{Introduction}
In our model Universe, the balance between expansion and collapse
stipulates that the size and the mass of a galaxy are set by its
formation epoch (see e.g. \citealt{press74}). Once most of the
galactic contents are in place, subsequent matter infall adds
little to the final mass budget (see e.g. \citealt{zemp13}). The total mass is dominated by dark matter; even though gas and stars might extend as
far, their densities drop faster with radius and therefore contribute
little to the integral over the virial volume. However, despite amounting to only $1$ per cent of the total galaxy luminosity or $<0.01$ per cent of the total mass, the stellar halo allows us to gauge the details of the mass distribution beyond the edge of the disk. 

The stars in the halo are more than mere tracers of the
potential. The dark matter radial density profiles are universal (and
hence featureless beyond the scale radius), or at least they appear to
be so for a considerable range of distances explored in numerical
simulations (e.g. \citealt{nfw97}). However, due to the plummeting star-formation efficiency in low-mass sub-halos (e.g. \citealt{bullock00}; \citealt{somerville02}), the
stellar halo formation is a much more stochastic process. The
lumpier accretion, combined with extremely long mixing times ($>1$
Gyr) can lead to a greater variety of stellar halo radial density
profiles (see e.g. \citealt{libeskind11}). Therefore, there is hope that by studying the phase-space
and chemical properties of halo stars today, we can uncover the fossil
record of the Milky Way's accretion history.

In order to quantify the stellar halo distribution, we often fit model profiles, such as power-laws and Einasto profiles (\citealt{einasto89}), to the stellar number counts. This approach has been widely used in the literature, and although these models may not represent a truly physical representation of the stellar halo, they provide a useful framework that can be compared with predictions from numerical simulations. Early work limited to Galactocentric radii $r \sim 20-30$ kpc found that the Milky Way stellar halo follows an oblate, single power-law distribution with minor-to-major axis ratio $q \sim 0.5-0.8$, and power-law index $\alpha \sim 2-4$ (e.g., \citealt{preston91}; \citealt{robin00}; \citealt{yanny00}; \citealt{newberg06}; \citealt{juric08}). More recent work, probing to greater distances in the halo, found evidence for a ``break'' in the stellar density profile at $r \sim 20-30$ kpc\footnote{In fact, the first hint of a break in the stellar halo density profile at $r \sim 25$ kpc was reported by \cite{saha85}, using a sample of $N \sim 29$ RR lyrae stars}. These studies find a power-law slope of $\alpha \sim 2-3$ can describe the stellar halo within $r \sim 20-30$ kpc, but a steeper slope with $\alpha \sim 3.8-5$ is required at larger distances (\citealt{bell08}; \citealt{watkins09}; \citealt{sesar11}; \citealt{deason11}). 

\cite{deason13a} argued that this broken profile could be caused by
the build-up of stars at their apocenters, either from the accretion
of one massive dwarf, or from several dwarfs accreted at a similar
epoch. On the other hand, \cite{beers12} claim that the change in
power-law slope near the break radius is caused by a transition from
an ``inner'' to an ``outer'' stellar halo population. Several groups
have found evidence for correlations between metallicity and
kinematics of halo stars, which perhaps suggest two distinct
populations (e.g. \citealt{carollo07}; \citealt{carollo10};
\citealt{nissen10}; \citealt{deason11a}; \citealt{hattori13};
\citealt{kafle13}). However, at present it is not obvious whether
these signatures can be produced purely from the accretion of dwarf
galaxies, or if some of these findings are biased by distance
uncertainites and/or contamination (e.g. \citealp{schonrich11, schonrich14}; \citealt{fermani13}).

It is clear that from a relatively ``simple'' measure of star counts, we can learn a great deal about the formation mechanism and/or past accretion history of the stellar halo. This bodes well for studies of stellar halos beyond the local group, where we are already able to measure the surface brightness profiles of these incredibly diffuse halos out to projected radii of $R \sim 50-70$ kpc (e.g. \citealt{radburn11}; \citealt{monachesi13}; \citealt{greggio14}; \citealt{vandokkum14}). The surface brightness profile of our nearest neighbor, M31, has now been mapped out to an impressive $R \sim 200$ kpc (e.g. \citealt{gilbert12}; \citealt{ibata14}). In contrast to our own Galaxy, these studies find no evidence for a break in the stellar density profile, and the star counts can be well-described by a single power-law with slope $\alpha \sim 3-3.5$. The differences between the stellar halo density profiles can give us an important insight into the contrasting accretion histories of the Milky Way and M31 (see \citealt{deason13a}).

Superimposed on the ``field'' (or phase-mixed) stellar halo distribution is a wealth of un-relaxed substructure in the form of streams, clouds and other overdensities (e.g. \citealt{ibata95}; \citealt{newberg02}; \citealt{belokurov06}; \citealt{belokurov07}; \citealt{juric08}). Most striking is the vast stream of tidal debris associated with the disrupting Sagittarius dwarf (\citealt{belokurov06}), where we are privy to a front-seat view of accretion in action. The presence of these recent accretion relics can significantly bias stellar halo number counts. Several studies have attempted to excise these known substructures, and only model the relatively phase-mixed halo component. However, it is important to understand the affect that these structures have on density profile measurements, especially when making comparisons with numerical simulations or stellar halos of external galaxies. For the latter, it is generally unfeasible to isolate the sort of substructures that we are able to identify in our own Galaxy.

Halo stars provide one of the best tracers of the Milky Way mass at large radii. \cite{deason12} compiled a sample of stellar halo stars with measured line-of-sight (LOS) velocities out to $r \sim 150$ kpc, and found a dramatic drop in LOS velocity dispersion beyond 50 kpc (see also \citealt{battaglia05}). If this drop reflects a fall in the circular velocity of the halo, then the Milky Way mass is likely below $\sim 10^{12}M_\odot$. This result agrees with several other stellar dynamical studies, which favor relatively low halo masses (e.g. \citealt{xue08}; \citealt{bovy12}). By contrast, other methods for estimating the Milky Way mass give larger values. For example, \cite{sohn13} recently used multi-epoch Hubble Space Telescope (\textit{HST}) images to measure the proper motion of the distant Milky Way satellite galaxy Leo I. Comparison of the large observed velocity to numerical simulations implies that the Milky Way mass is likely well above $10^{12} M_\odot$ (\citealt{boylan13}) . More generally, attempts to measure the total mass of the Milky Way using satellite galaxies (e.g. \citealt{wilkinson99}; \citealt{watkins10}), the Magellanic Clouds (e.g. \citealt{kallivayalil13}), the local escape speed (e.g. \citealt{smith07}), the timing argument (e.g. \citealt{li08}; \citealt{vandermarel12}; \citealt{gonzalez13}), and the methods already mentioned, have been distressingly inconclusive with total masses in the range $0.5-3 \times 10^{12}M_\odot$.

Halo stars have tremendous potential for constraining the Milky Way mass, since LOS velocities have been measured for many of them, but to make progress the \textit{mass-anisotropy-density} degeneracy must be addressed. Our mass measures based on halo star kinematics are limited by the uncertainty in the tracer density profile and velocity anisotropy. These systematic uncertainties are significant, and mass-measures can vary by up to factors of $\sim 5$ because of unknown tracer properties. Fortunately, the upcoming \textit{Gaia} mission and deep, multi-epoch \textit{HST} proper motion measurements (\citealt{deason13b}; \citealt{hstpromo}), will provide the missing transverse velocity information needed to measure the velocity ellipsoid of distant halo stars. However, we still have very little knowledge of the tracer density profile beyond 50 kpc. Thus, somewhat ironically, the ``simple'' task of counting stars will likely be the main bottleneck for dynamical mass measures of the Milky Way in the near future.

In this study, we use A-type halo stars selected from Sloan Digital Sky Survey (SDSS) data release 9 (DR9) photometry to measure the stellar halo density slope beyond $\sim 50$ kpc. These A-type stars comprise of blue horizontal branch (BHB) and blue straggler (BS) populations. The former stellar population, constitute our prime halo tracers and can probe out to $\sim 100$ kpc in the magnitude range used in this work ($18.5 < g < 20.5$). Our method models both BHB and BS populations simultaneously using photometric data alone, and includes the contribution from contaminants, such as QSOs. The combination of the large SDSS sky coverage ($\sim 14,000$ deg$^2$) and the accurate distance estimates provided by the BHB stars, allows us, for the first time, to constrain the outer density profile slope of the Milky Way stellar halo.

The paper is arranged as follows. In \S2.1  we describe the 
SDSS DR9 photometric data and our selection criteria for A-type 
stars. The remainder of \S2 describes our A-type star models and the absolute magnitude-color relations for the two populations. In \S3 we address the contribution of contaminants and the affects of photometric uncertainties on our modeling procedure. In \S4, we describe our likelihood-based method to determine the density profile of the stellar halo and in \S5 we present our results. Finally, we discuss the implications for the accretion history and the mass of the Milky Way in \S6, and summarize our main conclusions in \S7.

\section{A-type stars in SDSS data release 9}
\subsection{DR9 imaging}

\begin{figure*}
    \centering
    \includegraphics[width=18cm, height=4.5cm]{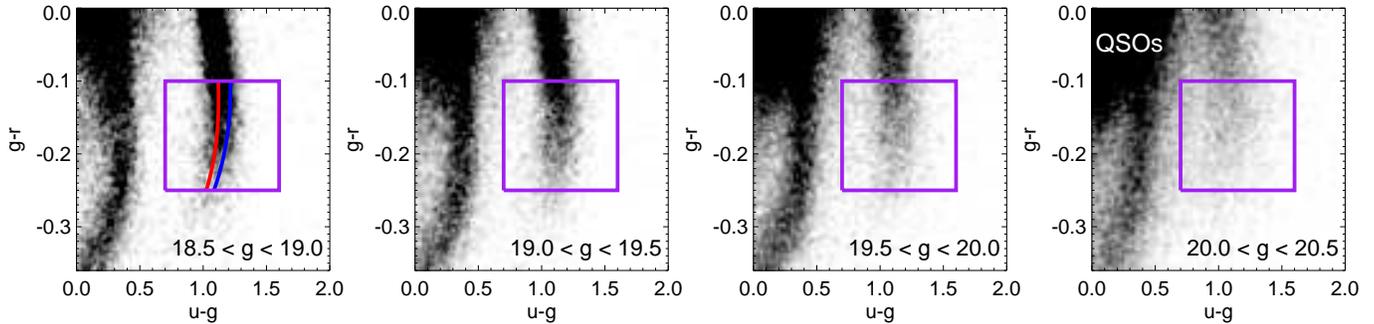}
    \caption[]{\small Color-color plots of high latitude ($|b| > 30^\circ$) stars selected from SDSS DR9 with $g$-band magnitudes in the range $18.5 < g < 20.5$. The ``claw'' sequence at $u-g \sim 1$ are BHB and BS A-type stars. The approximate ``ridgelines'' of these two populations are shown by the red (BSs) and blue (BHBs) lines for the brightest magnitude bin (see eqn. \ref{eq:ridge}). The purple box indicates the selection region for (blue) A-type stars in this work (see eqn. \ref{eq:ugr_sel}).  Each panel shows a different magnitude bin (increasing from left to right). QSOs and white dwarfs (WDs) populate the bluer $u-g$ region; WDs have a relatively tight sequence in $u-g$, $g-r$ space, while QSOs have a much broader distribution. At fainter magnitudes the A-type star claw becomes more blurred, and the QSO distribution (and to a lesser extent WDs - see Fig. \ref{fig:ug_contam}) starts to influence the region in color-color space where we select A-type stars.}
   \label{fig:ugr}
\end{figure*}

\begin{figure}
    \centering
    \includegraphics[width=8.5cm, height=6.8cm]{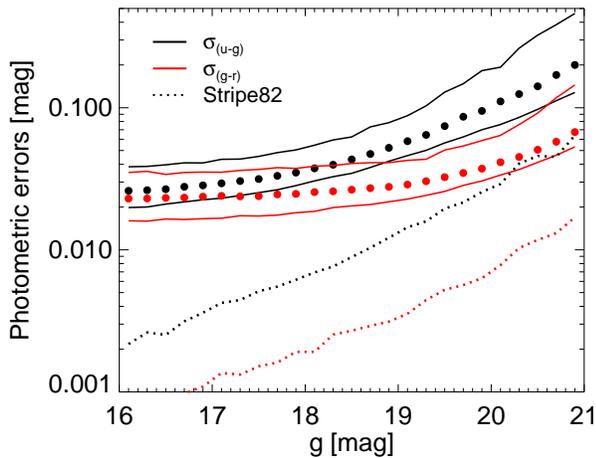}
    \caption[]{\small Photometric uncertainties in $u-g$ (black) and $g-r$ (red) as a function of $g$-band magnitude. The filled circles indicate the median values and the lines show the 5th and 95th percentiles. For comparison, the dotted lines show the (median) uncertainties for Stripe 82 photometry. The photometric errors for the Stripe 82 photometry are significantly lower than the single epoch SDSS measurements. In the magnitude range under consideration ($18.5 < g < 20.5$) the $u$-band errors are significantly increased relative to the brighter magnitudes used in DBE11 ($16 < g < 18.5$).}
   \label{fig:ug_err}
\end{figure}

The SDSS (\citealt{york00}) is an imaging and spectroscopic survey
covering over one quarter of the sky. Images are obtained simultaneously in five broad optical bands ($ugriz$; \citealt{fukugita96}) using a CCD camera (\citealt{gunn98}) on a 2.5-m telescope (\citealt{gunn06}) at Apache Point Observatory, New Mexico. Photometric and astrometric properties are derived through data processing pipelines developed throughout the course of the survey (\citealt{lupton01}; \citealt{smith02}; \citealt{stoughton02}; \citealt{pier03}; \citealt{ivezic04}; \citealt{tucker06}). The SDSS data release 9 (DR9; \citealt{sdss9}) provides the same sky coverage as its predecessor DR8 ($\sim 14 000$ deg$^2$), and contains all of the imaging data taken by the SDSS imaging camera\footnote{In the color and magnitude range under consideration, the SDSS DR9 photometric catalogs are almost identical to the previous DR8 release.}. We select high latitude ($|b| > 30$ deg) objects classified as stars by SDSS with clean $r$-band photometry. The magnitudes and colors we use in the following sections have been corrected for extinction following the prescription of \cite{schlegel98}.

In this study, we use BHB and BS A-type stars to map the density profile of the distant Milky Way stellar halo. These high latitude A-type stars are limited to a tight sequence in $u-g$, $g-r$ space (see text below and Fig. \ref{fig:ugr}) and suffer from relatively little contamination. This, in combination with the well-determined photometric parallaxes of BHB stars (see \S\ref{sec:absmag}), make these ideal tracers of the Milky Way halo. \cite{deason11} (hereafter, DBE11) used photometrically selected A-type stars from SDSS DR8, with $16.0 < g < 18.5$, to determine the stellar halo density profile within $r \sim 40-50$ kpc. Here, we extend this analysis to fainter magnitudes,  $18.5 < g < 20.5$, in order to measure the stellar density profile \textit{beyond 50 kpc}. In the magnitude range under consideration, BS/BHB stars have approximate heliocentric distances: $10 \lesssim D_{\rm BS}/\mathrm{kpc} \lesssim 75$,  $40 \lesssim D_{\rm BHB}/\mathrm{kpc} \lesssim 100$.

In Fig. \ref{fig:ugr} we indicate our A-type star selection in $u-g, g-r$ (or $ugr$ for short) space, $0.7 < u-g < 1.6$, $-0.25 < g-r < -0.1$, as a function of $g$-band magnitude. The range in $u-g$ is slightly expanded relative to DBE11 (cf. $0.9 < u-g < 1.4$) to account for the increased photometric errors at fainter magnitudes (see Fig \ref{fig:ug_err}). In addition, we impose a bluer $g-r$ selection ($g-r < -0.1$, DBE11 used $g-r < 0$) to minimize contributions from redder BS stars (see e.g. \citealt{brown10}) and QSOs. At fainter magnitudes the A-type ``claw'' sequence at $u-g \sim 1$ widens and the broad QSO distribution begins to impose on our A-type star selection box (see \S \ref{sec:qso_models}). 

\subsection{A-type star models}
\label{sec:at_models}

The modeling procedure used by DBE11 assumed that \textit{only} A-type stars (BHB or BS) were present in their sample, and they did not suffer from significant photometric errors. Our extension to fainter magnitudes means that we need to consider contamination from other populations, and the affect of photometric scattering; these model extensions are discussed in \S\ref{sec:contam} and \S\ref{sec:phot_scat}. 

Our A-type star models describe the distribution of BHB and BS stars in $ugr, m_g$ space. The 3D model for each stellar population depends on 1) the intrinsic $ugr$ color distribution, 2) the stellar density profile and 3) the absolute magnitude-($g-r$) color relation.

Owing to the higher surface gravity of BS stars, BHB and BS populations can be distinguished spectroscopically from their Balmer line profiles (e.g. \citealt{kinman94}; \citealt{clewley02}; \citealt{sirko04}; \citealt{xue08}; \citealt{deason12}). However, with photometry alone BHB and BS stars form distinct, but overlapping sequences in $u-g, g-r$ color-color space. DBE11 used bright A-type stars with available SDSS spectra to pinpoint the loci of BHB and BS populations in $u-g, g-r$ color-color space. These ``ridgelines'', i.e. the approximate centers of their distributions in $u-g$ as a function of $g-r$, are defined by third order polynomials:
\begin{eqnarray}
\label{eq:ridge}
(u-g)^{0}_{\rm BHB}&=& 1.167-0.775(g-r)-1.934(g-r)^2 \notag \\
&&+9.936(g-r)^3, \notag \\
(u-g)^0_{\rm BS}&=& 1.078-0.489(g-r)+0.556(g-r)^2 \notag \\
&&+13.444(g-r)^3, 
\end{eqnarray}
for $-0.25 < g-r < 0.0$. (see left-hand panel of Fig. \ref{fig:ugr}, and DBE11 Fig. 2)

The intrinsic spread of the two populations about their ridgelines are $\sigma_{\rm  BHB,0}(u-g)=0.04$ and $\sigma_{\rm  BS,0}(u-g)=0.045$. Gaussian distributions about these ridgelines are adopted to assign a membership probability based on $u-g$ color:

\begin{eqnarray}
\label{eq:prob0}
p(u-g|{\rm BHB}, g-r)\propto\mathrm{exp}\left(-\frac{\left[(u-g)-(u-g)_{\rm BHB}^0\right]^2}{2\sigma_{\rm BHB}^2}\right), \notag\\ 
p(u-g|{\rm BS}, g-r)\propto\mathrm{exp}\left(-\frac{\left[(u-g)-(u-g)_{\rm BS}^0\right]^2}{2\sigma_{\rm BS}^2}\right).
\end{eqnarray}
Here, the ridgelines (see eqn. \ref{eq:ridge}), $(u-g)^0$, depend on $g-r$ color.

The $g-r$ distribution of A-type stars depends on the stellar density profile, and the \textit{intrinsic} $g-r$ distribution of each population, i.e. the relative volume  densities of objects with different $g-r$. Thus, without \textit{a priori} knowledge of the stellar density profile, we cannot disentangle the intrinsic $g-r$ distribution. It is important that the apparent $g-r$ distribution for BS stars (and to a lesser extent BHB stars) at fixed $g$-band magnitude is different from the intrinsic one, and is a (strong) function of the density profile.

DBE11 estimated the relative fractions of BHB and BS stars in $g-r$ bins using the $u-g$ model distributions defined above.  However, in the fainter magnitude range under consideration here, the $u-g$ random errors are too large to be able to adequately distinguish between BHB and BS stars (see Fig. \ref{fig:ug_err} where $\sigma_{(u-g)} \sim 0.05-0.1$ between $18.5 < g < 20.5$). Instead, we use the stellar density profile derived by DBE11 for bright A-type stars ($16 < g < 18.5$), and the approximate numbers of BHB and BS stars in $g-r$ bins (see Table 1 in DBE11) to disentangle the intrinsic $g-r$ distributions of these populations. Note that this exercise assumes that the intrinsic distributions remain constant with magnitude.

\begin{figure}
    \centering
    \includegraphics[width=8.5cm, height=6.8cm]{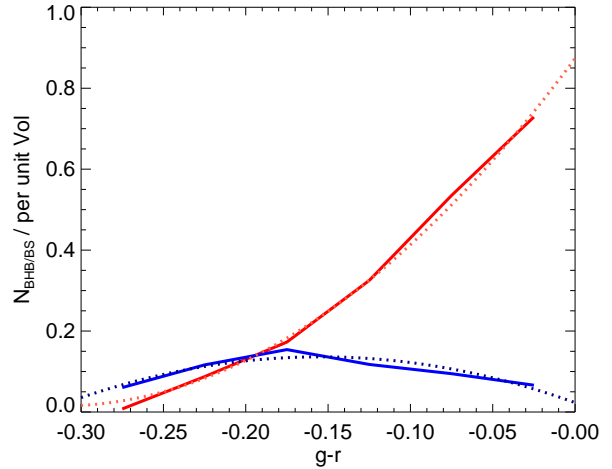}
    \caption[]{\small The intrinsic $g-r$ distribution of (bright) BHB (blue lines) and BS (red lines) stars. We show the number of stars per unit volume (in units of kpc$^{-3}$) as a function of $g-r$. The dotted lines indicate second order polynomial fits to the relations.}
   \label{fig:fgr_at}
\end{figure}

In Fig. \ref{fig:fgr_at} we show the resulting intrinsic $g-r$ distributions for (bright) BHB and BS stars. The BS stars have a steep $g-r$ distribution which rises sharply towards redder $g-r$ color, while the BHB stars are roughly constant with $g-r$. The dotted lines indicate polynomial fits to the $g-r$ distributions: $p(g-r | \mathrm{BHB}) \propto 0.70 -44.7(g-r)-144.7\left(g-r\right)^2$, $p(g-r | \mathrm{BS}) \propto 26.6+166.8(g-r)+265.7\left(g-r\right)^2$. Note that while the intrinsic $g-r$ distributions of BHB/BS stars are an important ingredient of the modeling procedure, our results are not significantly affected by our adopted parametrization. For example, our main conclusions are unchanged if we instead adopt a flat $g-r$ distribution for both populations.

Our A-type star membership probabilities can now be assigned based on $u-g, g-r$ colors:
\begin{eqnarray}
\label{eq:prob}
p(ugr|{\rm BHB})\propto p(g-r | \mathrm{BHB}) \, p(u-g | \mathrm{BHB}, g-r), \notag\\
p(ugr|{\rm BS})\propto p(g-r | \mathrm{BS}) \,  p(u-g | \mathrm{BS}, g-r).
\end{eqnarray}

\subsubsection{Absolute magnitude calibration}
\label{sec:absmag}
\begin{figure*}
  \begin{minipage}{0.5\linewidth}
    \centering
    \includegraphics[width=8.5cm, height=6.8cm]{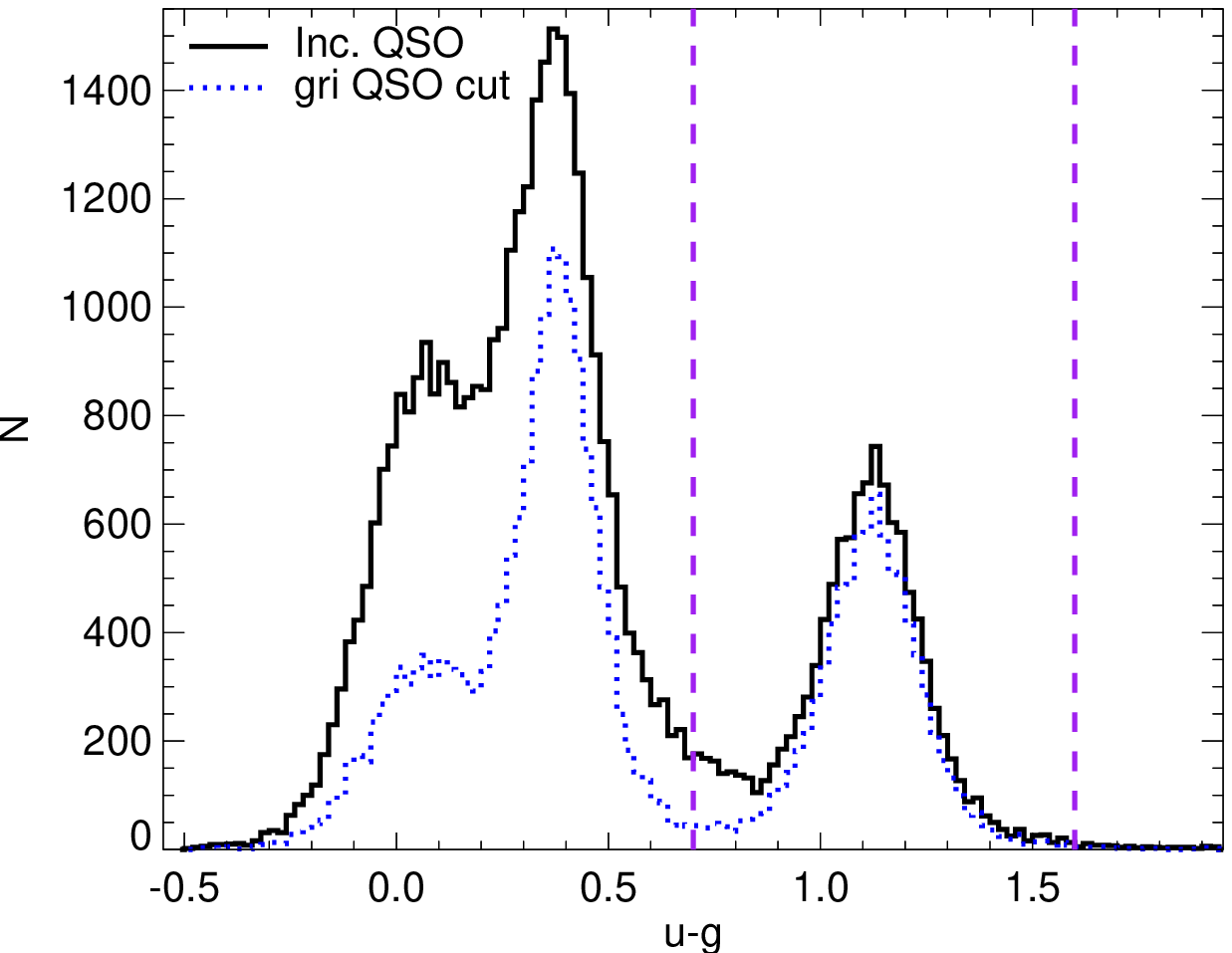}
   \end{minipage}
  \begin{minipage}{0.5\linewidth}
    \centering
   \includegraphics[width=8.5cm, height=6.8cm]{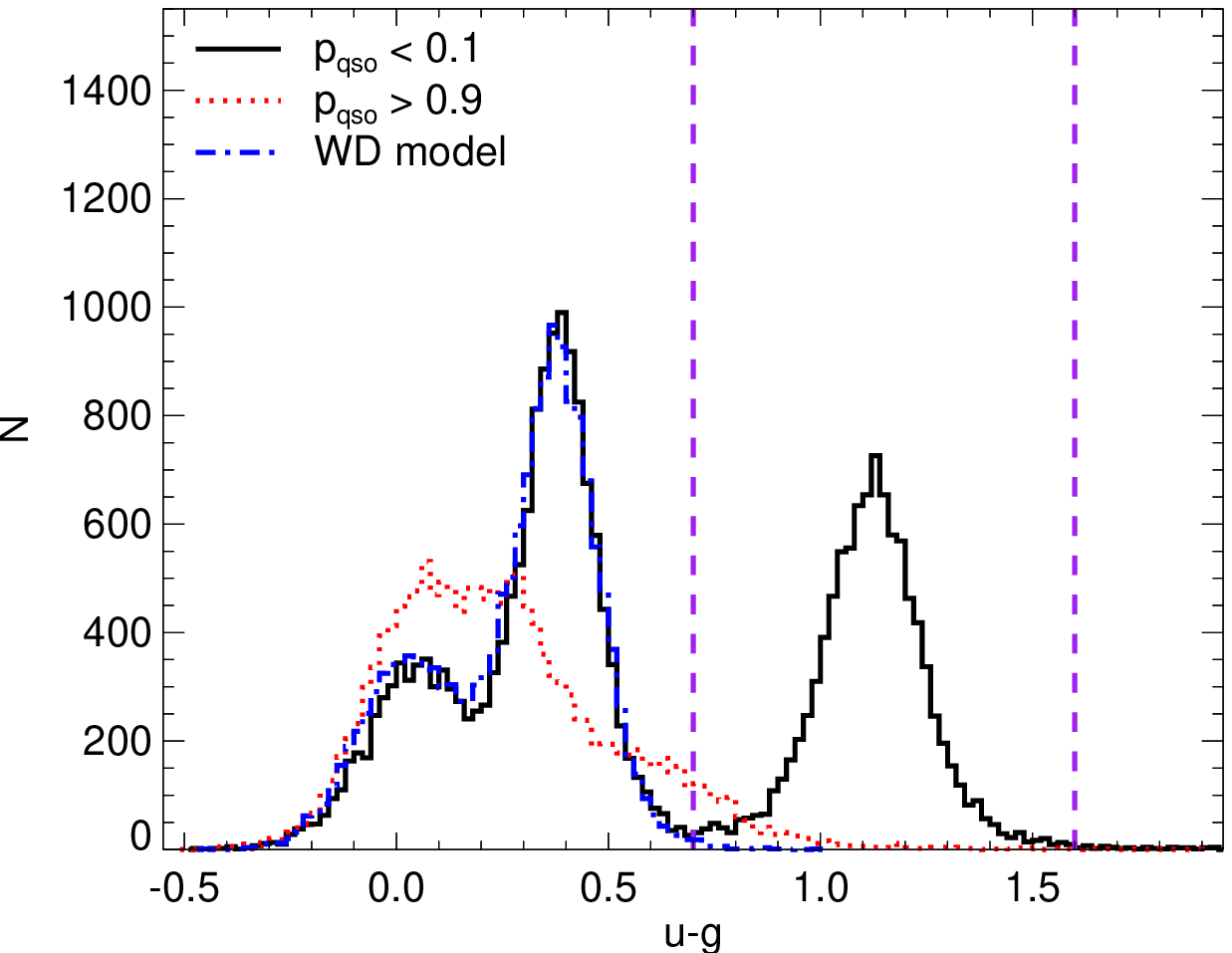}
   \end{minipage}
    \caption[]{\small \textit{Left panel:} The $u-g$ distribution of high latitude ($|b| > 30^\circ$) objects in the color and magnitude range $-0.25 < g-r < -0.1$ and $18.5 < g < 20.5$. The black line shows all objects, while the dotted blue line shows the distribution when a cut in $gri$ space to remove QSOs has been applied. Although this cut is not 100\% efficient and induces a bias against fainter objects, it illustrates the dominance of the QSO population as our main source of contamination. \textit{Right panel:} QSO probabilities are assigned using the XDQSO algorithm (see main text for details). High and low probability QSOs are shown with the dotted red and solid black lines respectively. The bluer low-probability QSOs are WD stars, a model for this population is shown with the dashed blue line (see Appendix \ref{sec:wd}). The contribution of WD stars in our A-type selection box (with $u-g > 0.7$) is minimal ($< 1\%$) and we do not consider them further in our analysis.}
   \label{fig:ug_contam}
\end{figure*}

BHB stars are intrinsically brighter than BS stars (by $\sim 2$ mag), and
their absolute magnitude varies little as a function of temperature or metallicity. By comparison, BS stars are intrinsically fainter and span a much wider range in absolute magnitude. We adopt the DBE11 absolute magnitude-color relations for these populations. DBE11 used star clusters with SDSS photometry published by \cite{an08} to calibrate the BHB absolute magnitudes, and stars selected in the Sagittarius stream with Stripe 82 photometry were used to calibrate the BS star absolute magnitudes (see \citealt{watkins09}). The derived relations are repeated here for completeness:
\begin{eqnarray}
\label{eq:absmag}
M_{g(\rm BHB)}&=& 0.434-0.169(g-r)+2.319(g-r)^2 \notag\\
&&+20.449(g-r)^3+94.517(g-r)^4, \notag\\
M_{g(\rm BS)}&=& 3.108+5.495(g-r),
\end{eqnarray}
where, $\sigma_{M_g(\rm BS)} \sim 0.5$. 

\section{Contamination and photometric scattering}

\subsection{Contamination}
\label{sec:contam}
Here, we consider the affect of contaminants on our A-type star selection box. In the left-hand panel of Fig. \ref{fig:ug_contam} we show the $u-g$ distribution for objects with $-0.25 < g-r < -0.1$ and $18.5 < g < 20.5$. The purple dashed lines indicate our $u-g$ selection boundary. QSOs and white dwarfs (WDs) have bluer $u-g$ colors than A-type stars, and these populations can clearly be seen for $u-g < 0.6$. Due to the relatively large photometric errors in the magnitude range under consideration, we must take into account these populations in our modeling procedure.

The blue-dotted line shows the $u-g$ distribution when a cut in $gri$ color space is applied to exclude QSOs (see \citealt{deason12} Fig.2). This cut is able to remove a significant amount of QSOs, but, owing to the large photometric errors at fainter magnitudes, it cannot remove all of them. Furthermore, this cut would also remove some A-type stars, especially those at fainter magnitudes with larger photometric errors. Thus, in order to avoid any biases, we do not apply this cut, but instead include a model for the QSO population in our analysis. The difference between the $u-g$ distributions with and without the $gri$ QSO cut illustrates that the QSOs constitute our main contaminant. In the following section we construct a model for the QSO population.

\subsubsection{QSO models}
\label{sec:qso_models}
\begin{figure*}
  \begin{minipage}{0.5\linewidth}
    \centering
    \includegraphics[width=8.5cm, height=6.8cm]{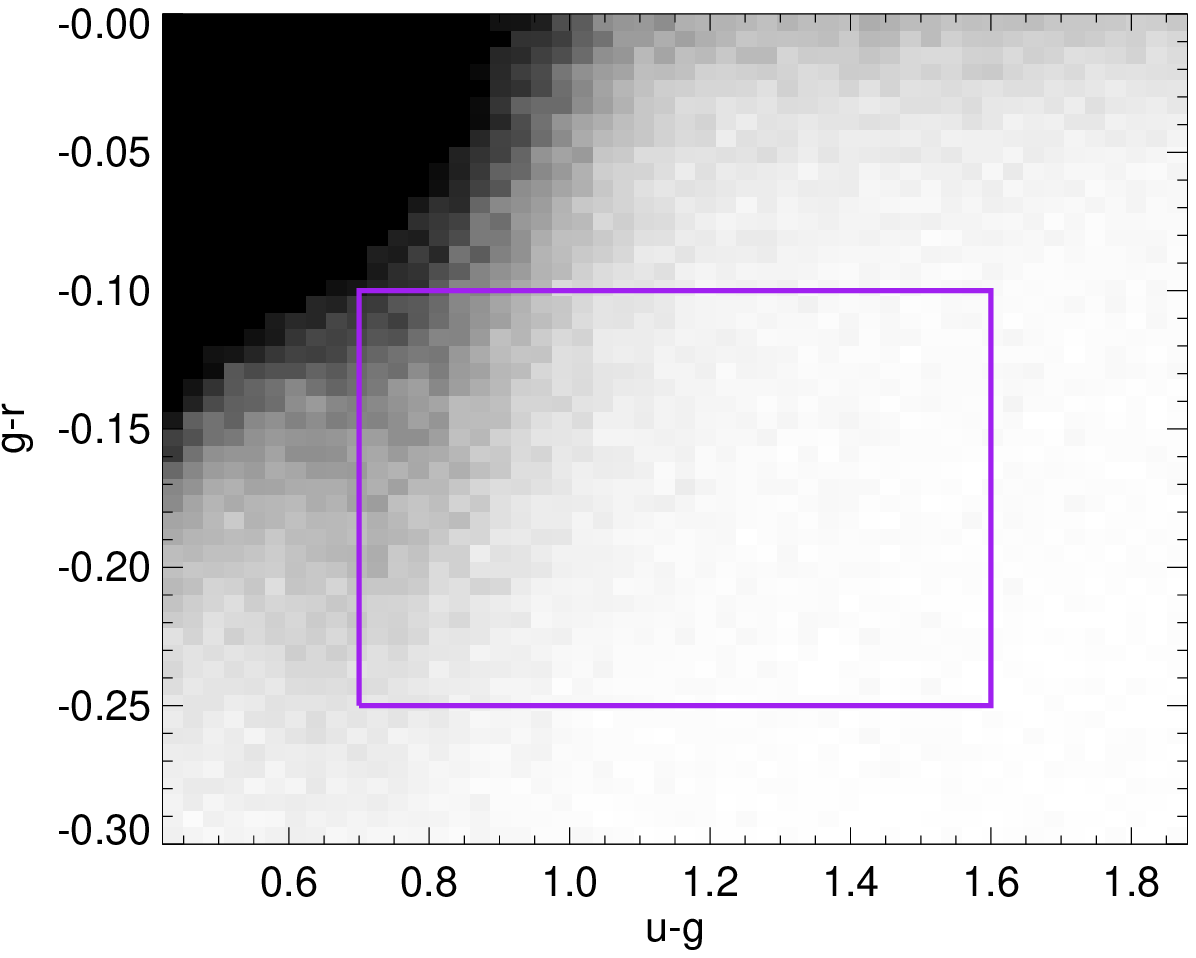}
   \end{minipage}
  \begin{minipage}{0.5\linewidth}
    \centering
   \includegraphics[width=8.5cm, height=6.8cm]{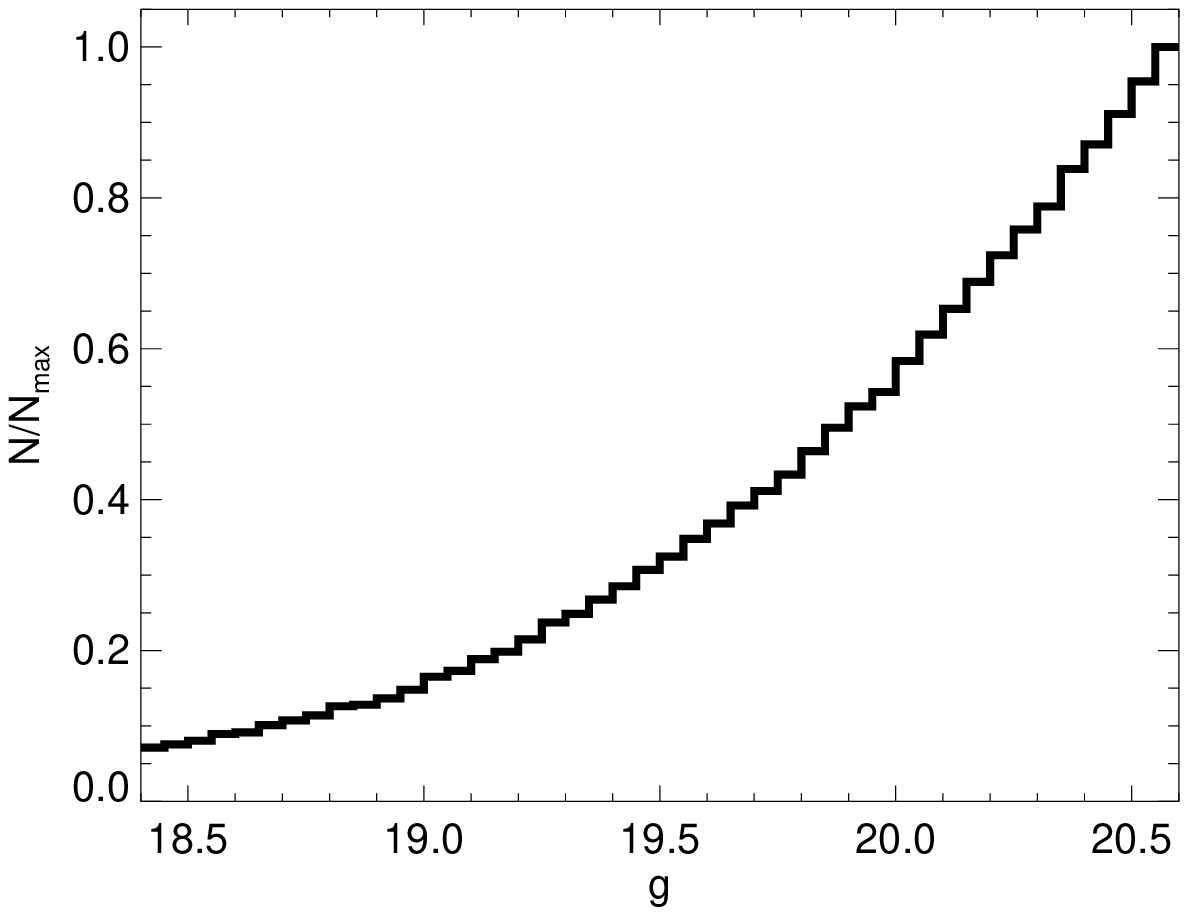}
   \end{minipage}
    \caption[]{\small \textit{Left panel:} The intrinsic $u-g$, $g-r$ distribution of the QSO model in the magnitude range $18.5 < g < 20.5$. The purple box indicates our $ugr$ selection for A-type stars. A significant number of QSOs can scatter into this box due to photometric errors. \textit{Right panel:} The $g$-band magnitude distribution for QSOs. The QSO luminosity function rises steeply with magnitude, thus the influence of QSOs becomes more important at fainter magnitudes.}
   \label{fig:qso_model}
\end{figure*}

To model the QSO population we make use of the XDQSO code\footnote{\url{http://www.sdss3.org/svn/repo/xdqso/tags/v0\_6/doc/build/}\\\url{html/index.html}} which is designed to calculate photometric QSO probabilities (see \citealt{bovy11}). The XDQSO algorithm was developed by \cite{bovy11} for efficient flux-based QSO target selection for SDSS data. Models of quasars in flux space were built by applying the extreme-deconvolution method (\citealt{bovy11b}) to spectroscopically confirmed QSOs in order to estimate the underlying density. This density is convolved with the flux uncertainties when evaluating the probability that an object is a QSO. We use these intrinsic photometric QSO models to construct the unconvolved (i.e. not convolved with photometric errors) QSO probability density function (PDF) in $ugr, m_g$ space.

The XDQSO code is applied to a uniform $ugriz$ distribution in the appropriate color and magnitude range. The QSO PDF is generated from the output likelihood distributions for low, medium and high redshift quasars:

\begin{equation}
P\mathrm{(QSO)}=\mathcal{L}_{z_{\rm low}}N_{z_{\rm low}}+\mathcal{L}_{z_{\rm mid}}N_{z_{\rm mid}}+\mathcal{L}_{z_{\rm high}} N_{z_{\rm high}}
\end{equation}
Here, $\mathcal{L}_{z_{\rm low}}$, $\mathcal{L}_{z_{\rm mid}}$ and $\mathcal{L}_{z_{\rm high}}$ are the relative flux likelihoods for low, medium and high redshift QSOs, and  $N_{z_{\rm low}}$, $N_{z_{\rm mid}}$ and $N_{z_{\rm high}}$ are the number counts at a given $i$-magnitude. The XDQSO algorithm computes the likelihoods and number counts for each QSO class given $ugriz$ fluxes. An acception-rejection algorithm is used to draw $u-g$, $g-r$ and $m_g$ values from this PDF, which gives the unconvolved QSO PDF in $ugr, m_g$ space:
\begin{equation}
\label{eq:qso_dens}
\nu_Q=p(ugr, m_g | \rm QSO)
\end{equation}
The resulting $u-g, g-r$ color distribution of the unconvolved QSO PDF is shown in the left-hand panel of Fig. \ref{fig:qso_model}. The purple box indicates our A-type star selection box. The inclusion of photometric errors scatters a significant number of QSOs into the box. The $g$-band magnitude distribution is shown in the right-hand panel. The QSO luminosity function increases steeply with magnitude, thus the QSO contribution becomes more significant at fainter magnitudes.

\subsubsection{White dwarf models}

We now consider the influence of the white dwarf (WD) population in our A-type star selection box. In the right-hand panel of Fig. \ref{fig:ug_contam} we show the $u-g$ distribution for low and high probability QSOs determined from the XDQSO algorithm. The low probability QSOs at bluer $u-g$ are WD stars. The blue dashed line shows the predicted $u-g$ distribution of our WD model (outlined in Appendix \ref{sec:wd}), where suitable photometric errors (see Fig. \ref{fig:ug_err}) have been included. Our WD model predicts a very small fraction of WD stars in our A-type star selection box ($< 1\%$). Thus, we can safely ignore the contribution of WD stars for the remainder of our analysis.

\subsection{Photometric scattering}
\begin{figure}
    \centering
    \includegraphics[width=8.5cm, height=5.7cm]{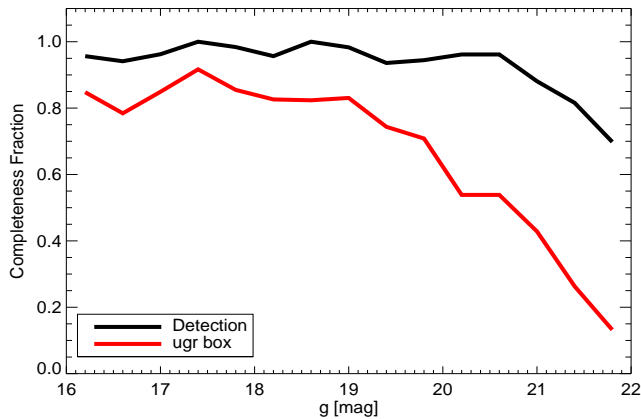}
    \caption[]{\small The fraction of A-type stars in Stripe 82 detected in our SDSS DR9 catalog as a function of magnitude. Based on detection alone we are $\sim 95\%$ complete over the relevant magnitude range. The red line shows the fraction when we also impose that the DR9 photometry is within the same $ugr$ bounds as the Stripe 82 stars. The fraction decreases due to photometric scattering, highlighting the importance of taking this effect into account.}
   \label{fig:photscat}
\end{figure}

\label{sec:phot_scat}
Our A-type star and QSO models describe the intrinsic, unconvolved populations. In the following section, we apply our likelihood method to the \textit{convolved} models, which takes into account the $ugr$ photometric uncertainties and their dependence on $g$-band magnitude. To illustrate the importance of taking photometric errors into account, we show in Fig. \ref{fig:photscat} an estimate of the completeness of our SDSS DR9 catalog. To estimate the completeness of our A-type sample we make use of the $\sim 2$ mag deeper stacked Stripe 82 photometry (\citealt{annis11}). Stars in the color range ($0.7 < u-g < 1.6$, $-0.25 < g-r < -0.1$) are selected from Stripe 82 and cross-matched with our DR9 catalog. The black line indicates the completeness fraction based on Stripe 82 stars which are detected in the DR9 catalog. The red line shows the completeness fraction when we also impose that the same stars fall within our $ugr$ selection box. Purely based on detection (i.e. the existence of an object with particular coordinates in both catalogs), our catalog is close to 95\% complete over the magnitude range under consideration ($18.5 < g < 20.5$). However, when we impose that the stars in Stripe 82 are also detected in the same $ugr$ range in the DR9 data, our recovered fraction is reduced, even at brighter magnitudes. This is due to photometric scattering; stars can be scattered into and out of our selection box and it is important to take this into account, especially at fainter magnitudes. 

The affect of photometric scattering can depend strongly on the shape of the PDF. For example, the intrinsic distribution of BS stars depends strongly on $g-r$ color, whereas BHB stars have a much weaker dependence on $g-r$ (see Fig. \ref{fig:fgr_at}). The steep gradient in $g-r$, means that a significant number of redder BS stars are likely to scatter into our selection box. Similarly, QSOs have a steep dependence on $g-r$ (see Fig. \ref{fig:qso_model}), which causes many to scatter into our A-type selection box at fainter magnitudes.

In the following section, our likelihood method is applied only to objects within our selection box. However, by constructing \textit{convolved} PDF models we are able to take into account the scattering of objects both into and out of this region in $ugr$ space. Our unconvolved models are also defined outside of the $ugr$ selection box (see eqn. \ref{eq:ugr_sel}). Thus, when these models are convolved with photometric uncertainties, the influence of populations which intrinsically lie outside of our $ugr$ selection box (i.e. redder BS stars, and QSOs), are considered in our models. In addition, the convolved models compensate for the affect of fainter stars ``leaking'' outside of the $ugr$ bounds.

\section{Likelihood Analysis}
\label{sec:ml}
Here, we describe our likelihood analysis used to model the density profile of distant halo stars. We apply our method to SDSS DR9 stars selected in the following magnitude and color range:
\begin{eqnarray}
\label{eq:ugr_sel}
18.5 <& g &< 20.5 \\ \notag
0.7 <& u-g &< 1.6\\ \notag
-0.25 <& g-r &< -0.1 
\end{eqnarray}

The number of stars of a particular population (i.e BHB or BS) in a given increment of magnitude and area on the sky is described by:
{\setlength\arraycolsep{0.1em}
\begin{eqnarray}
\label{eq:probglb}
\Delta N(m_g\!-\!M_g, \ell, b) &=& \rho(m_g\!-\!M_g,\ell, b) D^3(m_g\!-\!M_g)  \\
&& \quad \times \, \frac{1}{5}\mathrm{ln}10 \, \Delta m_g \, \mathrm{cos}b \, \Delta \ell \, \Delta b \notag
\end{eqnarray}
}

Here, we have used Galactic ($l, b$) coordinates and the heliocentric distance increment $\Delta D$ has been converted into the apparent magnitude increment via the relation $\Delta D=\frac{1}{5}\mathrm{ln}10 \, D \Delta m_g$. 

We combine equations (\ref{eq:prob}) and (\ref{eq:probglb}) to give the number of A-type stars in a cell of color, magnitude, and longitude and latitude space
{\setlength\arraycolsep{0.1em}
\begin{eqnarray}
 \Delta N_{\rm A} &=& p(ugr|{\rm BHB})\Delta N_{\rm BHB}+p(ugr|{\rm
   BS})\Delta N_{\rm BS} \\
&=&  N_{\rm tot}f_{\rm A} \nu_* (ugr, m_g, l,b) \Delta \underline{\mathbf{x}} \notag 
\end{eqnarray}

where $\Delta \underline{\mathbf{x}}= \mathrm{cos}b \Delta (u-g) \Delta (g-r)
 \Delta m_g \Delta \ell \Delta b$ is the volume element, and the stellar probability density is

{\setlength\arraycolsep{0.05em}
\begin{multline}
\label{eq:adens}
  \nu_* (ugr, m_g, l,b) = \frac{1}{5} \mathrm{ln}10 \, \times\\
 \left[f_{\rm BHB} \, p(ugr|{\rm BHB}) \, \rho_{\rm BHB}(m_g\!-\!M_g, \ell, b) D^3_{\rm BHB}(m_g\!-\!M_g) \right.
  \\ 
 \left. + f_{\rm BS} \, p(ugr|{\rm BS}) \, \rho_{\rm BS}(m_g\!-\!M_g, \ell, b) D^3_{\rm BS}(m_g\!-\!M_g) \right]
\end{multline}
}

Here, the absolute magnitudes of the BHB and BS populations depend on $g-r$ color ($M_g=M_g(g-r)$, see eq. \ref{eq:absmag}) and $f_{\rm BHB}$ and $f_{\rm BS}$ are constants used to ensure that the total number of BHB and BS stars equals the total number of A-type stars ($N_A=N_{\rm tot}f_A$). We take into account the uncertainty in the BS  absolute magnitudes by convolving the number density with a Gaussian magnitude distribution. This distribution is centered on the estimated absolute magnitude ($M^{\rm BS}_g=M^{\rm BS}_g(g-r)$) and has a standard deviation of $\sigma_{M_g}=0.5$.

In our analysis we assume the objects are A-type BHB and BS stars or QSOs. $N_{\rm tot}=N_{\rm A}+N_{\rm Q}$ where $N_{\rm A}=f_{\rm A}N_{\rm tot}=N_{\rm BHB}+N_{\rm BS}$ and $N_{\rm Q}=f_{\rm Q}N_{\rm tot}$.

The number of QSOs in a cell of color, magnitude, and longitude and latitude space is
{\setlength\arraycolsep{0.1em}
\begin{eqnarray}
 \Delta N_{\rm Q} &=&  N_{\rm tot} f_{\rm Q}\nu_{\rm Q} (ugr, m_g, \ell, b) \Delta \underline{\mathbf{x}}, 
\end{eqnarray}
}
where the QSO probability density, $\nu_Q$ is given by eqn. \ref{eq:qso_dens}.

The combined PDF for A-type stars and QSOs is then:
{\setlength\arraycolsep{0.1em}
\begin{eqnarray}
\Delta N_{\rm tot}&=&\Delta N_{\rm A}+\Delta N_{\rm Q} \\ \notag
&=&N_{\rm tot}\left((1-f_{\rm Q})\nu_*+f_{\rm Q}\nu_Q\right) \Delta \underline{\mathbf{x}}
\end{eqnarray}
}
Here, we have defined the unconvolved number densities for A-type stars and QSOs. These PDFs are convolved with the $ugr, m_g$ error distributions to give the convolved probability density distribution:
{\setlength\arraycolsep{0.1em}
\begin{eqnarray}
\tilde{\nu}&=&\nu \ast G \\
&=&\int\int\int \nu(ugr^*, m^*_{g}) \times \notag\\
&& \: \: \: \: \: \: \: \: \: \: \: \: \: \: G(ugr-ugr^*, m_g-m^*_g) \mathrm{d}(u-g) \, \mathrm{d}(g-r) \, \mathrm{d}m_g \notag
\end{eqnarray}
}
where, $G(ugr, m_g)$ is a 3D normal distribution in $u-g$, $g-r$ and $m_g$. The convolved densities are normalized in $ugr$, $m_g$, $\ell$ and $b$ space over the color and magnitude ranges specified in eqn \ref{eq:ugr_sel}, and over the area of the SDSS DR9 footprint.

The log-likelihood function can then be constructed from the convolved probability density distribution,
\begin{eqnarray}
  \mathrm{log}\mathcal{L}=\sum_{i=1}^{N_{\rm tot}}  \mathrm{log} \, \Big[ \big\{ \left( 1-f_{\rm Q} \right) \tilde{\nu}_* \left( ugr_i, m_{g,i}, \ell_i,b_i \right) &&  \notag \\ 
 +f_{\rm Q}\tilde{\nu}_{\rm Q} \left( ugr_i, m_{g,i}, \ell_i,b_i  \right) \big\}~\mathrm{cos}b_i \Big]. 
\end{eqnarray}

The overall fraction of QSOs, $f_{\rm Q}=1-f_{\rm A}$, and relative fraction of BHB stars, $f_{\rm BHB}$, are computed iteratively for each set of model parameters from the posterior PDFs:
\begin{eqnarray}
p(\mathrm{QSO}|ugr,m_g)=\frac{N_{\rm Q} \tilde{\nu}_Q}{N_{\rm Q} \tilde{\nu}_Q +N_{\rm A} \tilde{\nu}_*}\\ 
p(\mathrm{BHB}|ugr,m_g)=\frac{N_{\rm BHB} \tilde{\nu}_{\rm BHB}}{N_{\rm BHB} \tilde{\nu}_{\rm BHB} +N_{\rm BS} \tilde{\nu}_{\rm BS}}
\end{eqnarray}
These fractions give the relative contributions of BHBs, BSs and QSOs in our color-color, magnitude selection box, and ensure the contributions sum to give the total number of stars used in the modeling.

In the following section we outline our model assumptions for the stellar halo density profile. The inner stellar halo density profile ($r \lesssim 40$ kpc) is chosen based on constraints in the literature. We construct a marginal likelihood function by integrating over the adopted range of inner density profile parameters ($\alpha_1$, $\alpha_2$ and $r_c$, see eqns \ref{eq:model} and \ref{eq:inner_range}). The maximum likelihood parameters for the outer stellar halo profile ($r_b$ and $\alpha_{\rm out}$, see eqn. \ref{eq:model}) are found using a brute-force grid search.

\subsection{Model assumptions}

The likelihood method described above is general, and can be applied to any number of model density profiles. Here, we outline the model assumptions applied in our analysis.

The aim of this study is to quantify the outer stellar halo density fall-off. We assume BHB and BS stars follow the same density distribution, which we parametrize as a spherical triple power-law:
\begin{equation}
\label{eq:model}
\rho(r)_{\rm BHB, BS} \propto \begin{cases} r^{-\alpha_1} & r \le r_c  \\
                              r^{-\alpha_2} & r_c < r \le r_b \\ 
                              r^{-\alpha_{\rm out}} & r > r_b \\ 
\end{cases}
\end{equation}

A toy model of this power-law profile is shown in Fig. \ref{fig:toy} for illustration.
In this work, we only consider spherical radial profiles. It is well-known that the stellar halo is flattened in the inner regions (with minor-to-major axis ratio $q \sim 0.6-0.8$; e.g. DBE11; \citealt{sesar11}), but it is unlikely that such a flattened profile can exist to large radii. Here, we concentrate on the radial density fall-off, and defer a study of the variation of shape of the stellar halo with radius to future work. In Appendix \ref{sec:ftest} we create mock datasets which have flattened stellar halos in the inner regions, and discuss the implications of assuming sphericity at all radii in our modeling procedure.

Previous work has shown that within $r \approx 40-50$ kpc, the MW stellar density follows a broken power-law (e.g. \citealt{bell08}; \citealt{sesar11}; DBE11). Based on this past work, we assume the following constraints on $r_c$, $\alpha_1$ and $\alpha_2$:

\begin{eqnarray}
\label{eq:inner_range}
r_c &\in& [20, 30] \mathrm{kpc} \notag \\
\alpha_1 &=& 2.5 \notag \\
\alpha_2  &\in& [3.5, 5.0]
\end{eqnarray}

In our analysis, we marginalize over the inner profile parameters. This assumes flat priors over the parameter space given above. Note that the inner-most power-law slope is kept fixed as this has little affect on the outer-most power-law ($\alpha_{\rm out}$). The free parameters in our analysis are thus, $r_b$ and $\alpha_{\rm out}$, and we consider values in the range: $r_b \in [30, 70] \mathrm{kpc}$ and $\alpha_{\rm out} \in [2.0, 10.0]$.

\begin{figure}
  \centering
  \includegraphics[width=8.5cm, height=6.8cm]{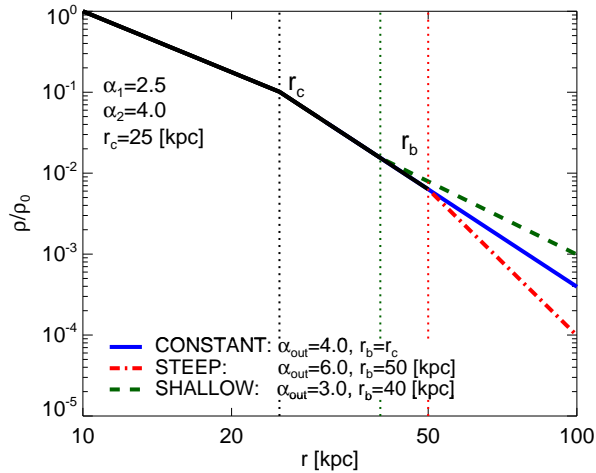}
  \caption[]{\small A toy model of our adopted density profile. The ``steep'' model is similar to the profile we measure for the Milky Way stellar halo (see results in \S\ref{sec:results}), while the ``shallow'' and ``constant'' models more closely resemble the M31 stellar halo (see discussion in \S\ref{sec:m31})}
 \label{fig:toy}
\end{figure}

\begin{figure*}
  \centering
  \includegraphics[width=16cm, height=8cm]{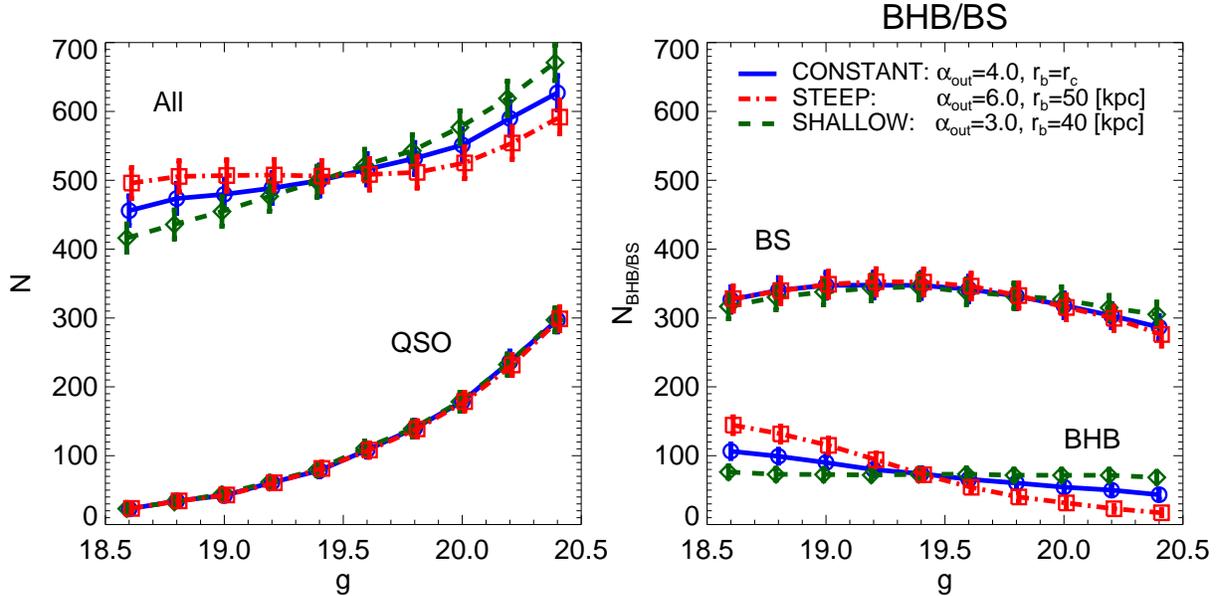}
  \caption[]{\small The magnitude distributions of the mock data. In the left panel we show the overall distribution (inc. BHBs, BSs and QSOs), and the contribution from QSO contaminants. The right panel shows the magnitude distributions for both BHB and BS populations.}
 \label{fig:gmag_fake}
\end{figure*}

\subsection{Tests with mock data}
\label{sec:mock}

\begin{figure*}
  \begin{minipage}{0.5\linewidth}
    \centering
    \includegraphics[width=8.5cm, height=6.8cm]{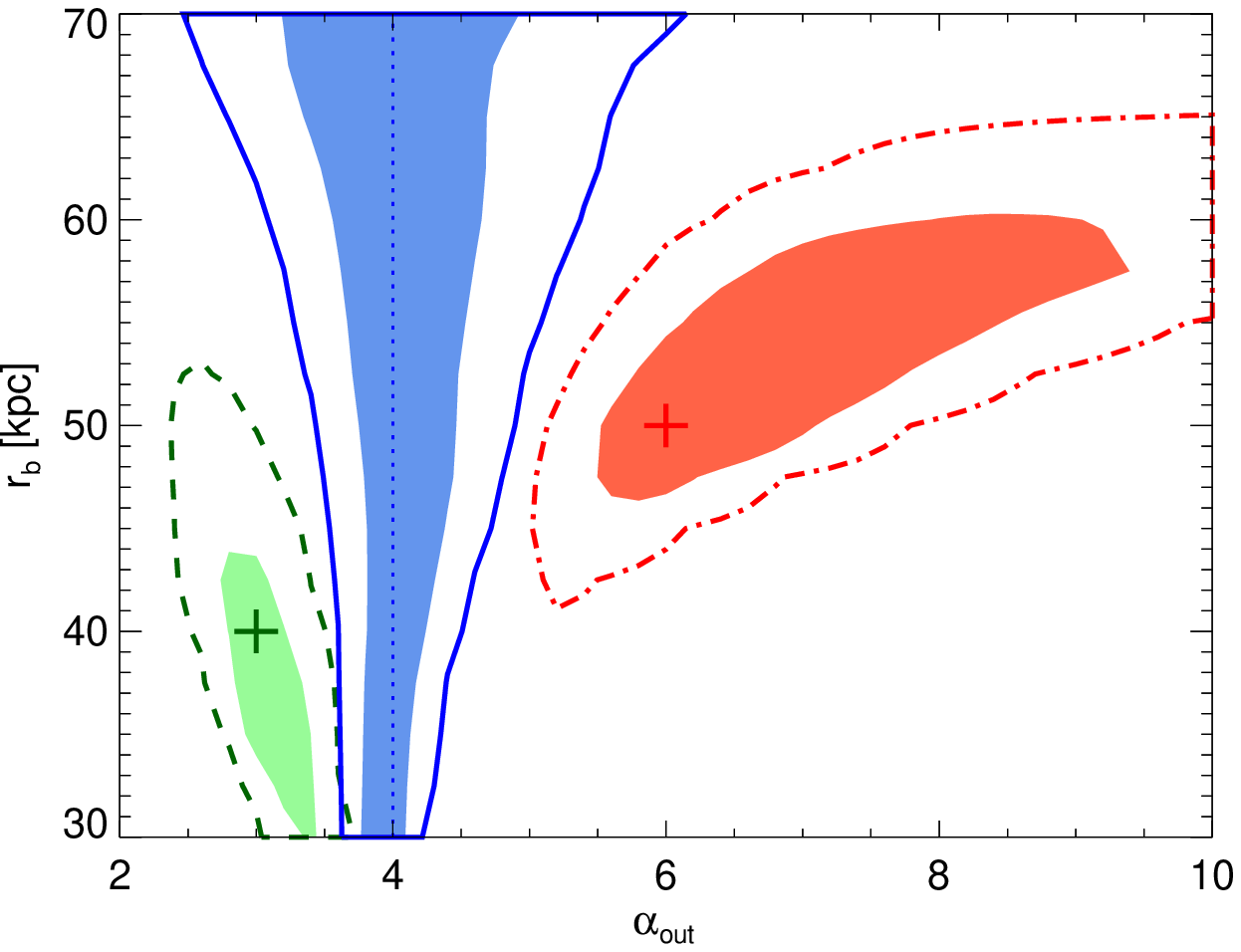}
   \end{minipage}
  \begin{minipage}{0.5\linewidth}
    \centering
   \includegraphics[width=8.5cm, height=6.8cm]{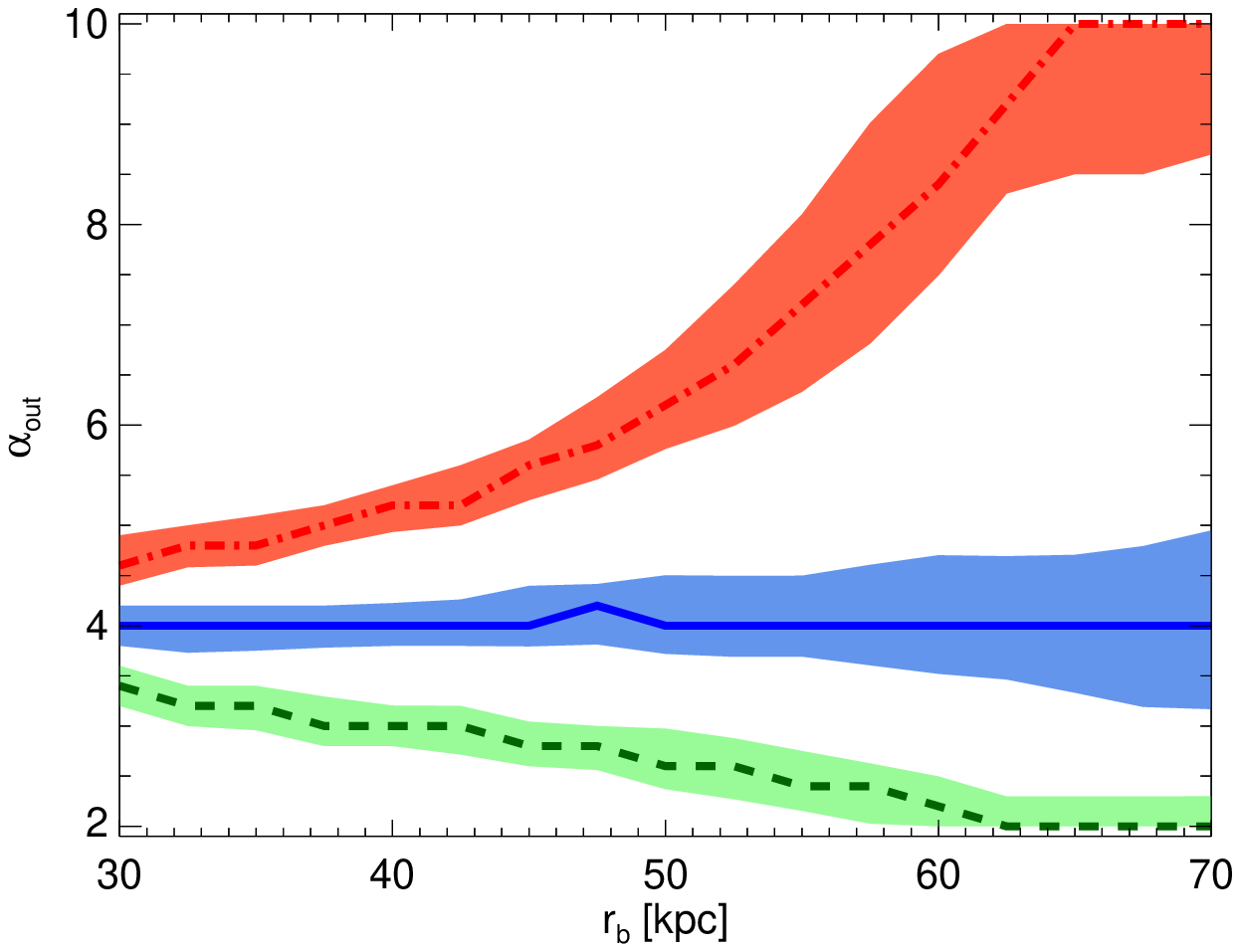}
   \end{minipage}
    \caption[]{\small Results of applying our likelihood analysis to mock data. \textit{Left panel:} Likelihood contours, where the filled and unfilled contours indicate the 1- and 2-$\sigma$ confidence regions, respectively. The green, blue and red contours indicate the shallow, constant and steep models, respectively. \textit{Right panel:} The maximum likelihood outer slope for different fixed values of $r_b$. The lines show the maximum likelihood parameters, and the shaded regions indicate the 1-$\sigma$ uncertainties.}
   \label{fig:like_fake}
\end{figure*}

To demonstrate the ability of our modeling technique, we apply our likelihood method to ``mock'' data. For our mock data, we assume a (fixed) inner profile with $r_c=25$ kpc, $\alpha_1=2.5$ and $\alpha_2=4.0$, and consider three different outer halo models: a ``shallow'' model with $r_b =40$ kpc, $\alpha_{\rm out}=3.0$, a ``constant'' fall-off model with $\alpha_{\rm out}=\alpha_2=4.0$, and a ``steep'' model with $\alpha_{\rm out}=6$, $r_b=50$ kpc. In all mock datasets, we adopt overall population fractions of $f_{\rm BHB}=0.18$ and $f_{\rm Q}=0.23$.

The following steps are applied to generate the mock data:

\begin{itemize}
\item A-type stars (BHB and BS) and QSOs are drawn from the (unconvolved) model PDFs defined in eqns \ref{eq:qso_dens} and \ref{eq:adens} using an acception-rejection algorithm. Objects are generated in $ugr, m_g,  \ell, b$ space from uniform color, magnitude distributions, and $l,b$ are drawn randomly from the surface of a sphere.
\item For the A-type stars, $g$-band magnitudes are converted to heliocentric distances using absolute-magnitude color relations appropriate for each stellar population. The BS absolute magnitudes are scattered about their mean relations assuming $\sigma_{M_g} \sim 0.5$.
\item Only high latitude objects (with  $|b| > 30^\circ$), inside of the SDSS DR9 footprint are considered. 
\item The 3D error distribution in $ugr, m_g$ space appropriate for our SDSS DR9 sample (see Fig. \ref{fig:ug_err}) is applied to the mock data. After photometric scattering, only objects lying within the bounds defined in eqn. \ref{eq:ugr_sel} are considered. 
\item Our mock data sets are generated with the same number of stars as our SDSS DR9 sample with known substructures removed ($N =5213$, see \S\ref{sec:subs})
\end{itemize}

The magnitude distributions of our three models (``shallow'': dashed green, ``constant'': solid blue, ``steep'': dot-dashed red) are shown in Fig. \ref{fig:gmag_fake}. The overall magnitude distributions for the three models are similar, but there are clear differences between the BHB star distributions; this is not surprising given that the distance range of the BS stars generally lie within the (fixed) inner density profile (e.g. approximately $~77\%$ of BS stars are within $r=30$ kpc).

The results of applying our likelihood method to the mock data are shown in Fig. \ref{fig:like_fake}. In the left-hand panel we show the likelihood contours in $\alpha_{\rm out}$, $r_b$ space. The filled and unfilled contours indicate 1- and 2-$\sigma$ confidence regions respectively. In all cases, our method is able to reproduce, within the uncertainties, the true density profiles. In the right-hand panel we show the maximum likelihood $\alpha_{\rm out}$ values for different fixed values of break radius $r_b$. The lines indicate the median values, and the shaded regions encompass the 1-$\sigma$ confidence regions. The ``steep'' models show a characteristic steepening as the adopted break radius is increased. Such models are clearly distinguished from shallower profiles.

In the above exercise we adopt the same inner profile as the input mock data ($\alpha_1=2.5, \alpha_2=4.0, r_b=25$ kpc). However, when we apply our method to the SDSS DR9 data we marginalize over a wide range of inner profile parameters (see eqn. \ref{eq:inner_range}). In Appendix \ref{sec:ftest} we show that this flexibility can compensate for biases induced by assuming an inaccurate inner stellar halo profile.

\subsection{Treatment of known substructures}
\label{sec:subs}

In our analysis, we consider the effect of known structures on our density profile estimates. There are two large structures in the regime of our sample which could affect our results: the Virgo overdensity (\citealt{juric08}) and the Sagittarius (SGR) stream (\citealt{ibata95}).

The Virgo overdensity is located at high latitudes and mainly affects the brightest BS stars in our sample (in the distance range $10 \lesssim D/\mathrm{kpc} \lesssim 20$). We isolate stars belonging to Virgo by applying the following cut in Galactic coordinates (see \citealt{bell08}; DBE11):

\begin{equation}
0 < X < 30, \quad\quad  X=63.63961\sqrt{2(1 - \sin b)}.
\end{equation}

Stars belonging to the SGR stream are present over the full magnitude range of our sample (see Fig. \ref{fig:dmm}, \citealt{belokurov14}). Thus, we locate possible SGR stars according to their position on the sky (see \cite{deason12} Fig. 7).

Finally, a small fraction of distant BHB stars in our sample coincide with two known dwarf galaxies; Sextans ($D \sim 90$ kpc) and Ursa Minor ($D \sim 60$ kpc). Stars in the regions of these dwarfs ($N \sim 200$) are excluded.  

Our selection of A-type stars gives $N_{\rm tot} = 10,787$ objects including SGR and Virgo, and $N_{\rm tot} = 5213$ when objects in the regions of these known overdensities are excluded. Below, we apply our likelihood analysis to our SDSS DR9 sample both with and without these large substructures.

\section{Results}
\label{sec:results}
\begin{table}
\caption{\small{Maximum Likelihood Results}\label{tab:like}}
\begin{center}
\renewcommand{\tabcolsep}{0.12cm}
\renewcommand{\arraystretch}{1.4}
\begin{tabular}{|l c  c c c c|}
    \hline 
     &  $r_{b}$ [kpc] & $\alpha_{\rm out}$ & $f_{Q}$ & $f_{\rm BHB}$ & $\Delta \mathrm{log}\mathcal{L}$\tablenotemark{a}
    \\
    \hline
 \multicolumn{2}{|l}{\textbf{Exc. Virgo \& SGR}} & & & &\\
    \hline
   & 30 & $4.8^{+0.3}_{-0.2}$&  0.234 & 0.193 & -4.9 \\
   & 35 & $4.8^{+0.2}_{-0.3}$&  0.233 & 0.192 &-4.5\\
   & 40 & $4.6^{+0.3}_{-0.2}$&  0.231 & 0.197 & -4.7\\
   & 45 & $4.8^{+0.5}_{-0.3}$&  0.232 & 0.194 &-5.3\\
   & 50 & $6.0^{+0.6}_{-0.9}$& 0.235 & 0.180 &-4.6\\
   & 55 & $7.0^{+0.8}_{-0.8}$&  0.235 & 0.178 &-2.4 \\
   & 60 & $8.4^{+1.2}_{-0.8}$&  0.236 & 0.178 &-0.6 \\
   & 65 & $>7.1$\tablenotemark{b}&   0.236 & 0.181 & 0.0\\
   & 70 & $>7.1$\tablenotemark{b}& 0.235 & 0.189 & -0.3\\
       \multicolumn{3}{|r}{$(r_b, \, \alpha_{\rm out})$\tablenotemark{c}$=( \, 65^{+5}_{-6} \, \mathrm{kpc}$, \,$>6.2$\tablenotemark{b} )} &&&\\
 \multicolumn{6}{|r|}{$(r_c, \, \alpha_{1}, \, \alpha_{2}, \,r_b, \, \alpha_{\rm out})_{\rm ML}$\tablenotemark{d} $= ( \, 25 \, \mathrm{kpc}, 2.5 \,, 4.5 \,, 65  \, \mathrm{kpc} \,, 10 \,$)} \\
       \hline
\multicolumn{2}{|l}{\textbf{Inc. Virgo \& SGR}} & & & &\\
    \hline
   & 30 & $4.4^{+0.2}_{-0.2}$ & 0.198 & 0.249 &-22.2 \\
   & 35 & $4.4^{+0.2}_{-0.3}$& 0.195 & 0.248 &  -22.8 \\
   & 40 & $4.6^{+0.2}_{-0.2}$& 0.194 & 0.245 &-20.6\\
   & 45 & $5.4^{+0.2}_{-0.3}$&  0.196 & 0.230 &-8.2\\
   & 50 & $6.0^{+0.5}_{-0.3}$& 0.195 & 0.228 & -2.2 \\
   & 55 & $7.2^{+0.5}_{-0.5}$&  0.196 & 0.223 & 0.0\\
   & 60 & $9.0^{+0.8}_{-0.8}$& 0.196 & 0.221 &-2.6 \\
   & 65 & $7.8^{+1.2}_{-0.8}$&  0.194 & 0.242 & -10.9\\
   & 70 & $8.4^{+1.0}_{-1.0}$&   0.193 & 0.248 & -14.3\\
     \multicolumn{3}{|r}{$(\,r_b, \, \alpha_{\rm out})$\tablenotemark{c}$=( 55^{+4}_{-3} \, \mathrm{kpc}, 7.2^{+1.6}_{-0.7}\,$)} &&&\\
     \multicolumn{6}{|r|}{$(r_c, \, \alpha_{1}, \, \alpha_{2}, \,r_b, \, \alpha_{\rm out})_{\rm ML}$\tablenotemark{d} $= ( \, 30 \, \mathrm{kpc}, 2.5 \,, 3.5 \,, 55  \, \mathrm{kpc} \,, 7.2 \,$)} \\
    \hline
  \end{tabular}
  \vspace{-0.5pt}
  \tablenotetext{a}{Difference in log-likelihood from maximum likelihood value.}
\tablenotetext{b}{2-$\sigma$ lower limits}
\tablenotetext{c}{Joint maximum likelihood result for $r_b$ and $\alpha_{\rm out}$ after marginalizing over the inner stellar halo parameters $r_c$ and $\alpha_2$.}
\tablenotetext{d}{The maximum likelihood parameters from the 4-D space of $r_c, \alpha_2, r_b, \alpha_{\rm out}$. Note, $\alpha_1$ is kept fixed (see eqn. \ref{eq:inner_range}).}
 \end{center}
\end{table}

In this section, we apply our likelihood technique to our sample of A-type stars selected from SDSS DR9. In Fig. \ref{fig:like} we show the likelihood results. In the left-hand panel, we show the 1- and 2-$\sigma$ confidence levels for the outer slope ($\alpha_{\rm out}$) and break radius ($r_b$). The dashed red and solid black lines show the results both with and without SGR and Virgo. In the right-hand panel we show the maximum likelihood outer slope for different fixed values of break radius.

There is a strikingly steep fall-off ($\alpha_{\rm out} \sim 6-10$) in the stellar halo density beyond $\sim 50-60$ kpc. This holds true even when the SGR and Virgo overdensities are included in the analysis. The implications of this result for the accretion history and mass profile of the Milky Way are discussed in \S\ref{sec:diss}. We note that the location of a ``break'' in the stellar density at $r \sim 50-60$ kpc is coincident with the apocenter of the SGR leading arm (see \citealt{belokurov14} Fig. 4). \cite{deason13a} showed that ``breaks'' in the stellar halo density are strongly linked to the apocenters of their accreted constituents; thus, the location of a break close to the apocenter of SGR agrees very well with this hypothesis.

In our analysis, the fraction of QSO contamination ($f_{\rm Q}$) and the overall BHB star fraction ($f_{\rm BHB}$) are found iteratively for each model PDF. In Fig. \ref{fig:fractions_contour} we show the variation of these fractions with our model parameters. Our maximum likelihood model parameters are summarized in Table \ref{tab:like}. Here, we also give the maximum likelihood parameters over the 4-D space ($r_c, \alpha_2, r_b, \alpha_{\rm out}$).  The maximum likelihood inner profile parameters ($r_c, \alpha_2$) listed here, dominate over the likelihood when we marginalize over a range of parameters (see eqn. \ref{eq:inner_range}).

\begin{figure*}
  \begin{minipage}{0.5\linewidth}
    \centering
    \includegraphics[width=8.5cm, height=8.5cm]{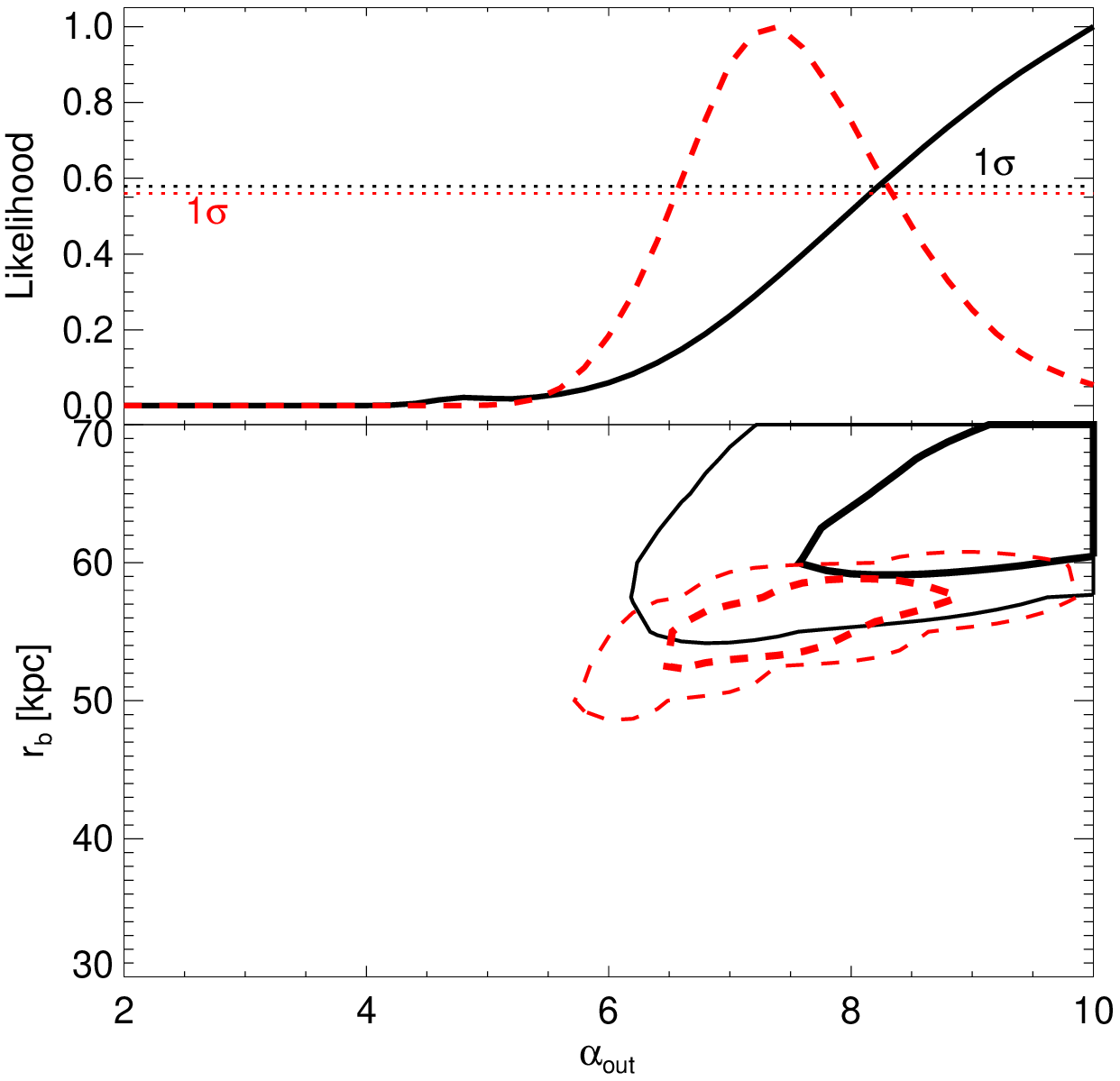}
   \end{minipage}
  \begin{minipage}{0.5\linewidth}
    \centering
   \includegraphics[width=8.5cm, height=8.5cm]{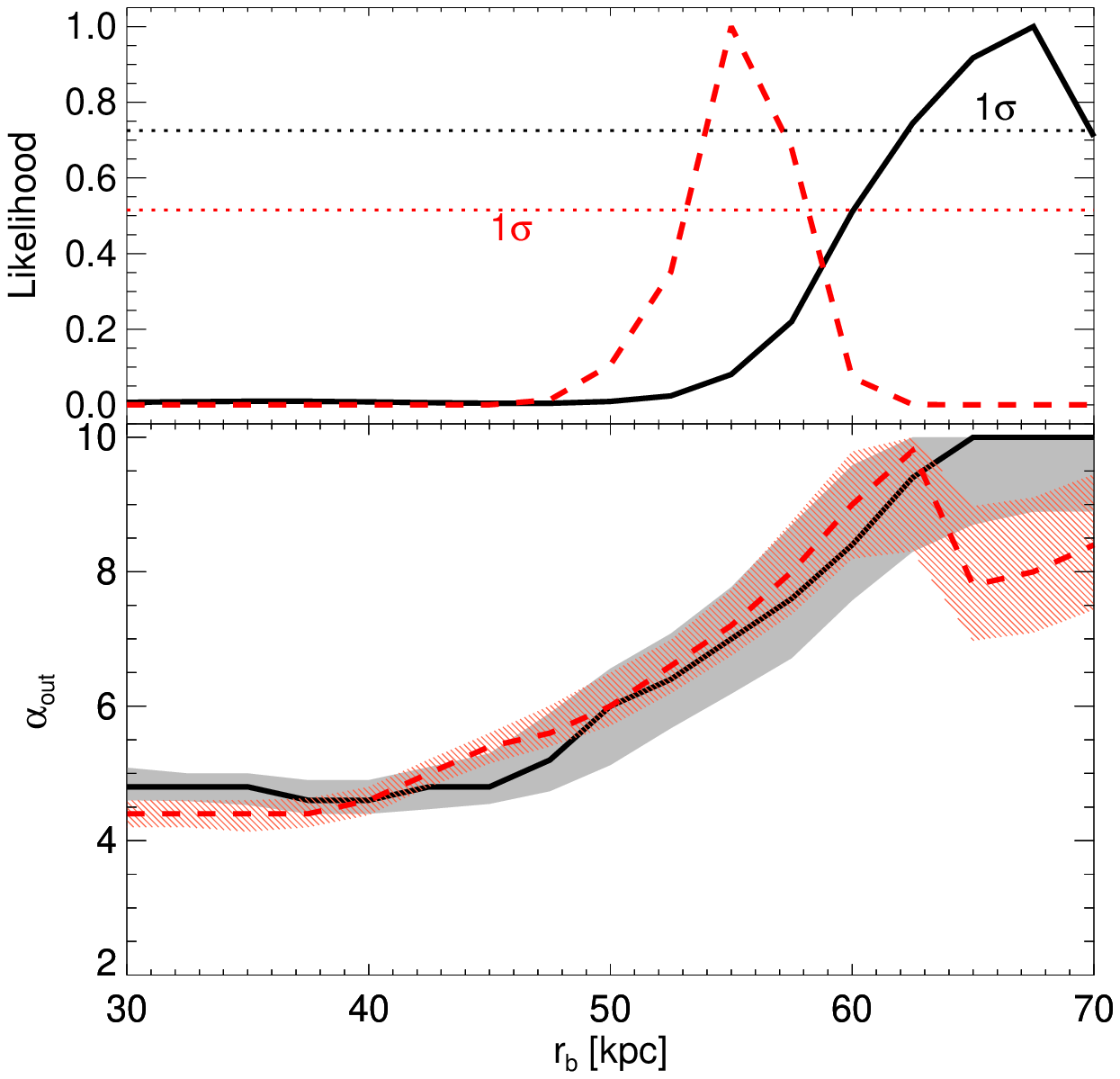}
   \end{minipage}
    \caption[]{\small Maximum likelihood results. \textit{Left panel:} The contours indicate the 1- and 2-$\sigma$ confidence levels respectively. The dashed red and solid black lines show the results both with and without SGR and Virgo. The top inset shows the marginalized likelihood distribution for the outer slope. The horizontal dotted lines indicate the 1-$\sigma$ confidence levels. \textit{Right panel:}  The maximum likelihood outer slope for different fixed values of $r_b$. The lines show the maximum likelihood parameters, and the shaded regions indicate the 1-$\sigma$ uncertainties. This plot illustrates the strong covariance between $r_b$ and $\alpha_{\rm out}$. The top inset shows the marginalized likelihood distribution for the break radius.}
   \label{fig:like}
\end{figure*}

\begin{figure}
    \centering
    \includegraphics[width=8.5cm, height=8.5cm]{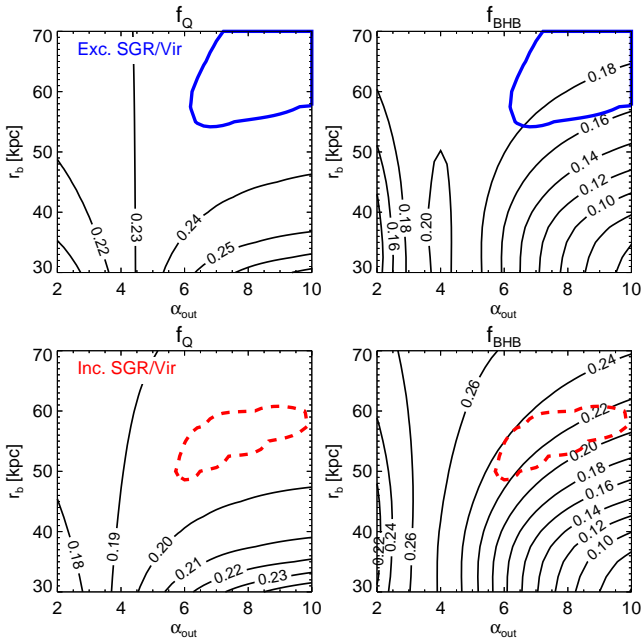}
    \caption[]{\small The variation in QSO fraction ($f_Q$, left panels) and BHB fraction ($f_{\rm BHB}$, right panels) with our model parameters. SGR and Vir are excluded/included in the top/bottom panels respectively. The 2-$\sigma$ confidence contours for the model parameters are also shown to highlight the high likelihood parameter space.}
   \label{fig:fractions_contour}
\end{figure}

\begin{figure*}
\begin{minipage}{0.5\linewidth}
  \centering
  \includegraphics[width=8.5cm, height=11.9cm]{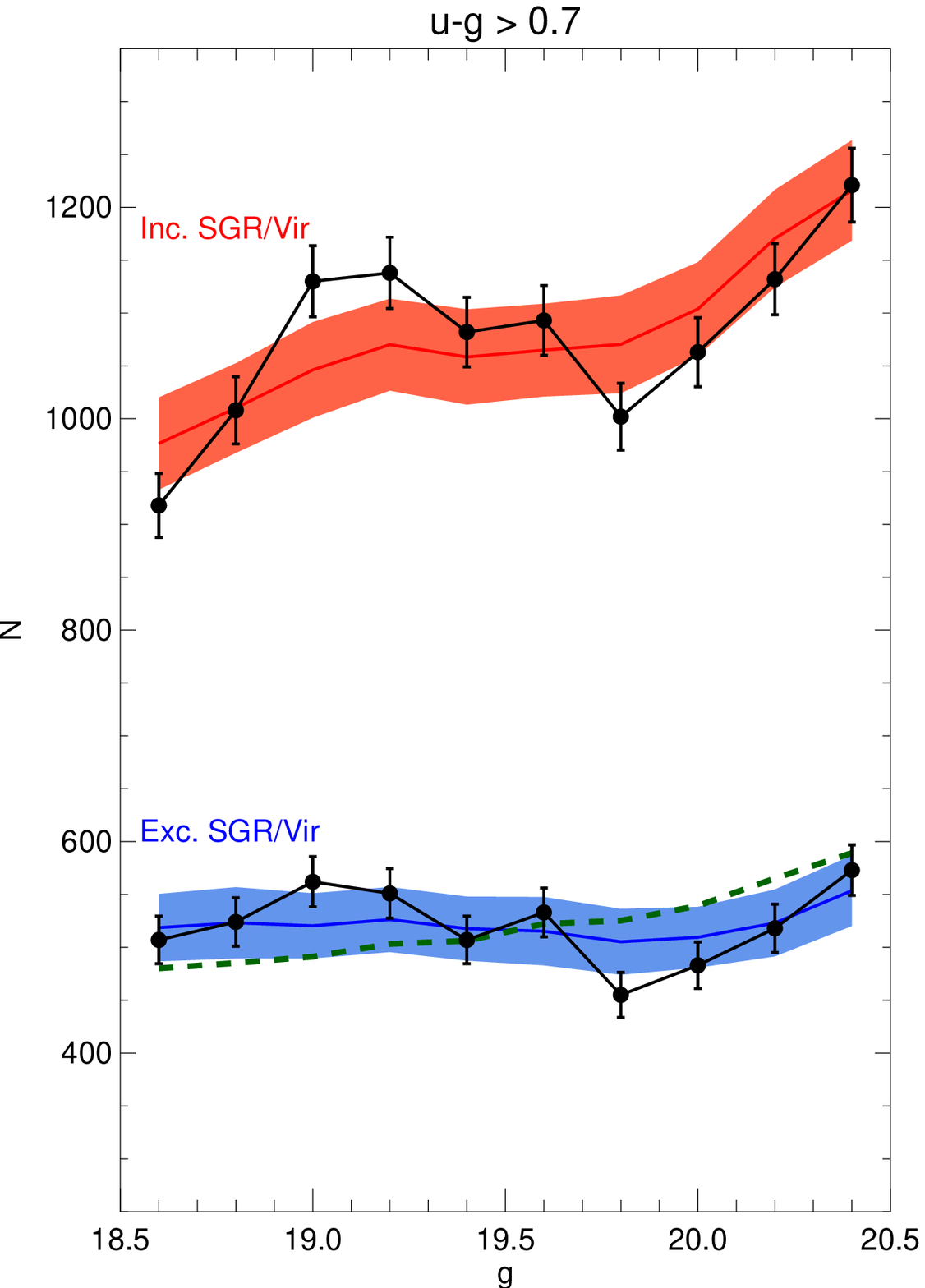}
 \end{minipage}
 \begin{minipage}{0.5\linewidth}
   \includegraphics[width=8.5cm, height=11.9cm]{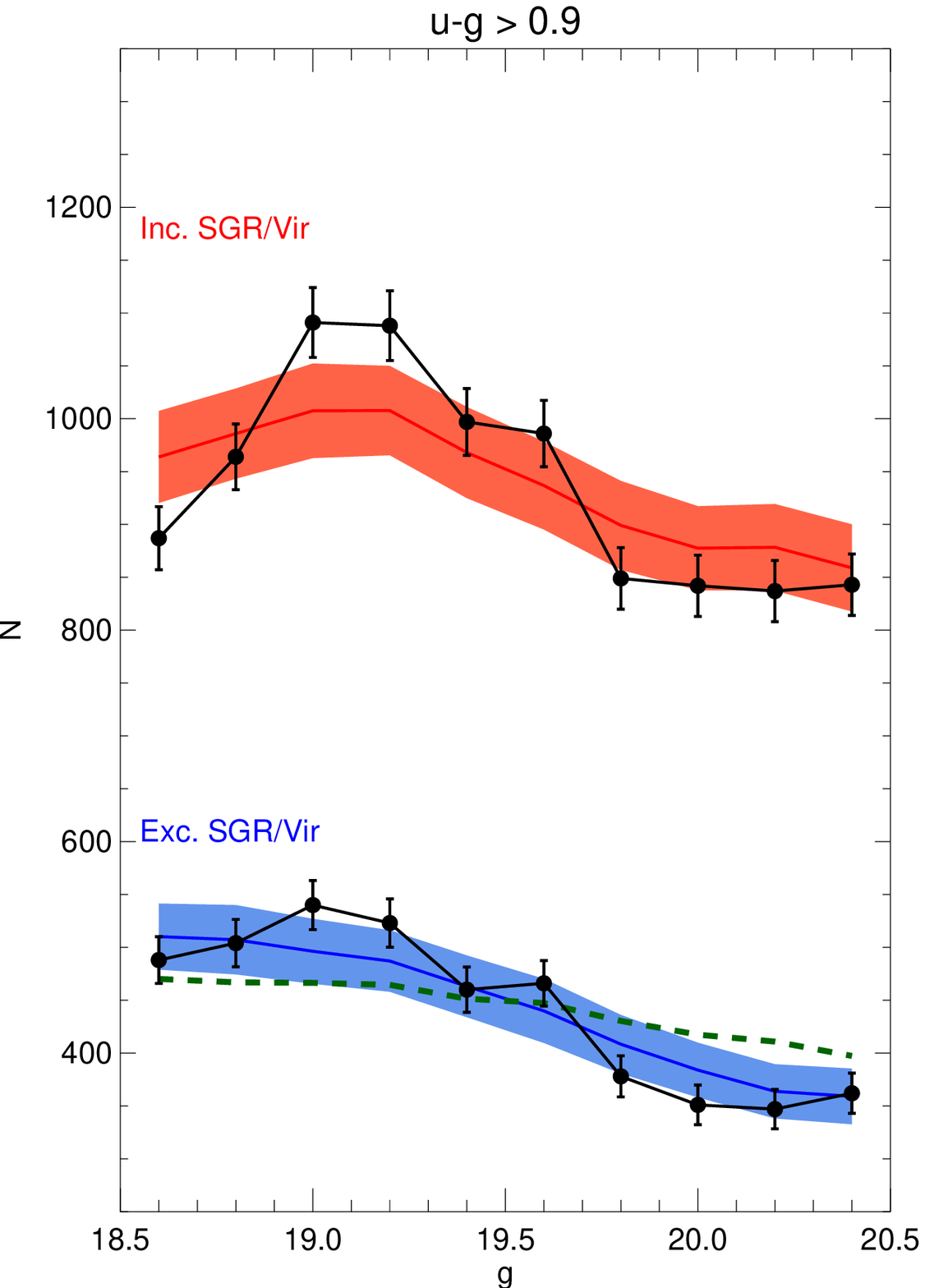}
 \end{minipage}
    \caption[]{\small The $g$-band magnitude distribution of our DR9 data sample. The data is shown with the black points and error bars. The best-fit models are shown by the shaded red (including SGR/Virgo) and blue (excluding SGR/Virgo) regions. The error bars indicate the model uncertainties due to Poisson noise. In the left panel, all stars in our $ugr$ selection box are shown ($u-g > 0.7$), and in the right panel we only show stars with $u-g > 0.9$ to reduce the QSO contribution. For comparison, the median relation for a (less likely) model with a shallow outer slope ($\alpha_{\rm out}[r_b =50$ kpc$]=3.5$), is shown with the dashed-green line.}
   \label{fig:gmag}
\end{figure*}

\begin{figure*}
\begin{minipage}{0.5\linewidth}
  \centering
  \includegraphics[width=8.5cm, height=4.25cm]{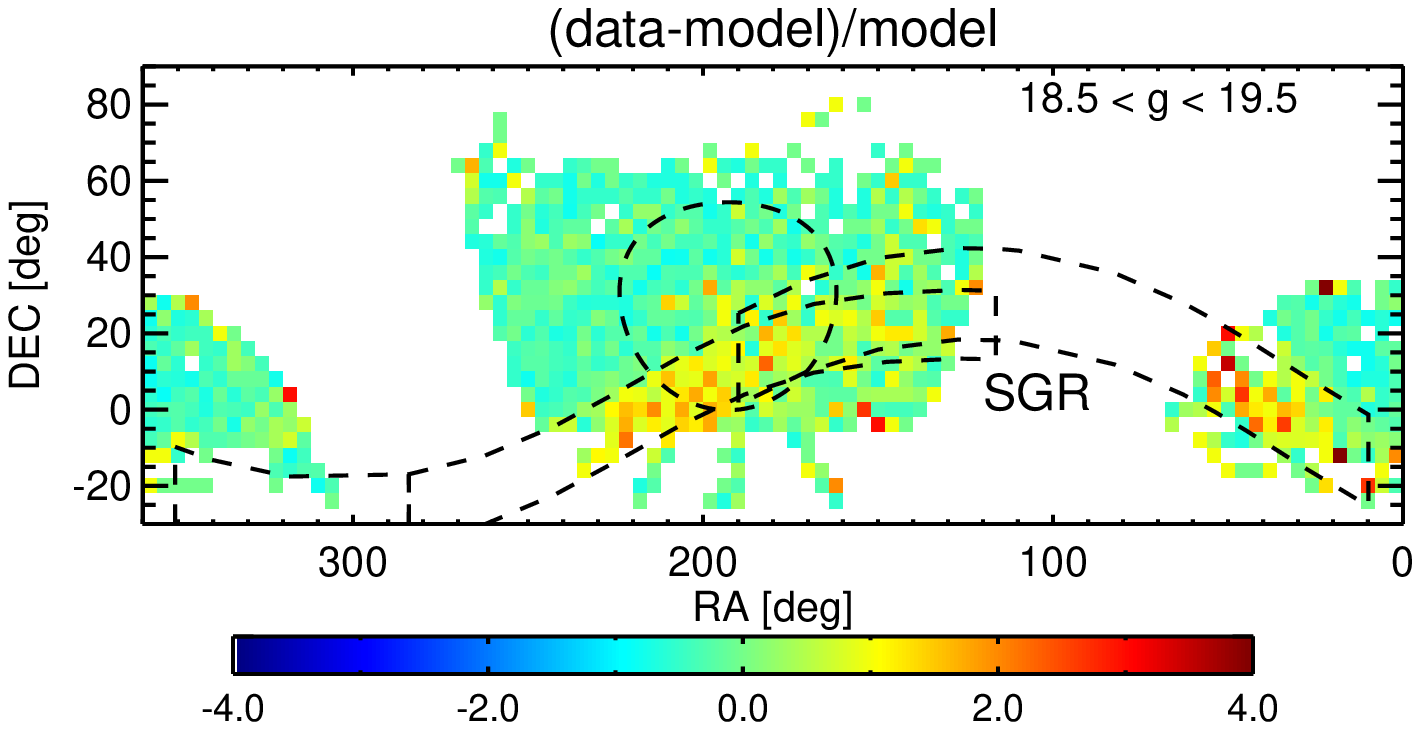}
 \end{minipage}
 \begin{minipage}{0.5\linewidth}
   \includegraphics[width=8.5cm, height=4.25cm]{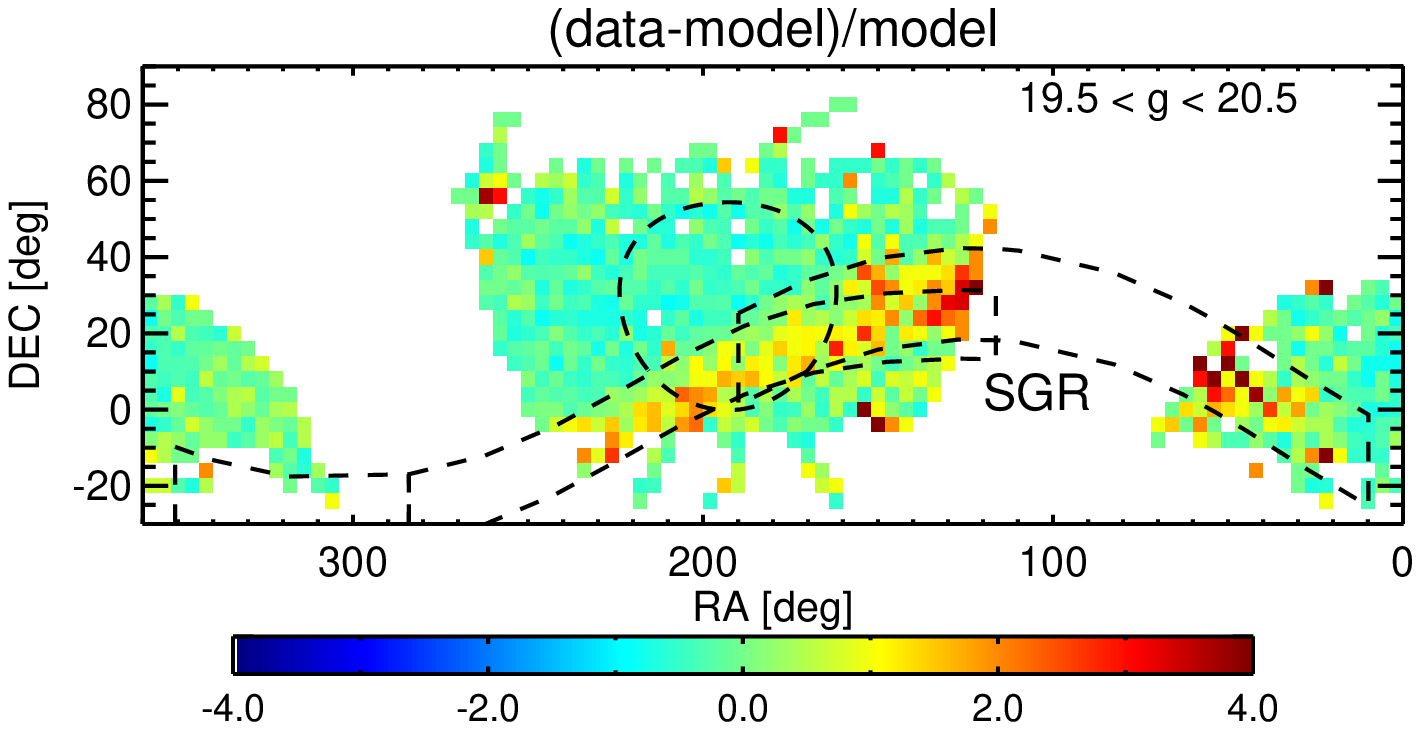}
 \end{minipage}
 \caption[]{\small Data minus model residuals on the sky for our best-fit model (excluding SGR and Vir in the modeling) in Equatorial coordinates. The two panels are split into ``bright'' ($18.5 < g < 19.5$; left panel) and ``faint'' ($19.5 < g < 20.5$; right panel) magnitude bins. Sagittarius dominates the overdense regions, but away from the stream the residuals are close to zero.}
 \label{fig:dmm}
\end{figure*}

In Fig. \ref{fig:gmag} we show the $g$-band magnitude distribution of the data and best-fit model. The blue/red shaded regions show the best-fit models when SGR and Virgo are excluded/included respectively. In the left-panel we show all the stars in our $ugr$ selection box. In the right-panel we only show stars with $u-g > 0.9$ to reduce the influence of QSOs on the $g$-band magnitude distribution. There is good agreement between the data and models, especially when known overdensities are excluded. Finally, we show in Fig. \ref{fig:dmm} the residuals of our models and data on the sky (in equatorial coordinates). We split into two magnitude bins; ``bright'' ($18.5 < g < 19.5$) and ``faint'' ($19.5 < g < 20.5$).  The increasing dominance of SGR is evident in these two panels.  However, we note that away from the SGR stream, the residuals are close to zero. 

\section{Discussion}
\label{sec:diss}
\begin{figure*}
  \begin{minipage}{0.5\linewidth}
    \centering
    \includegraphics[width=8.5cm, height=6.8cm]{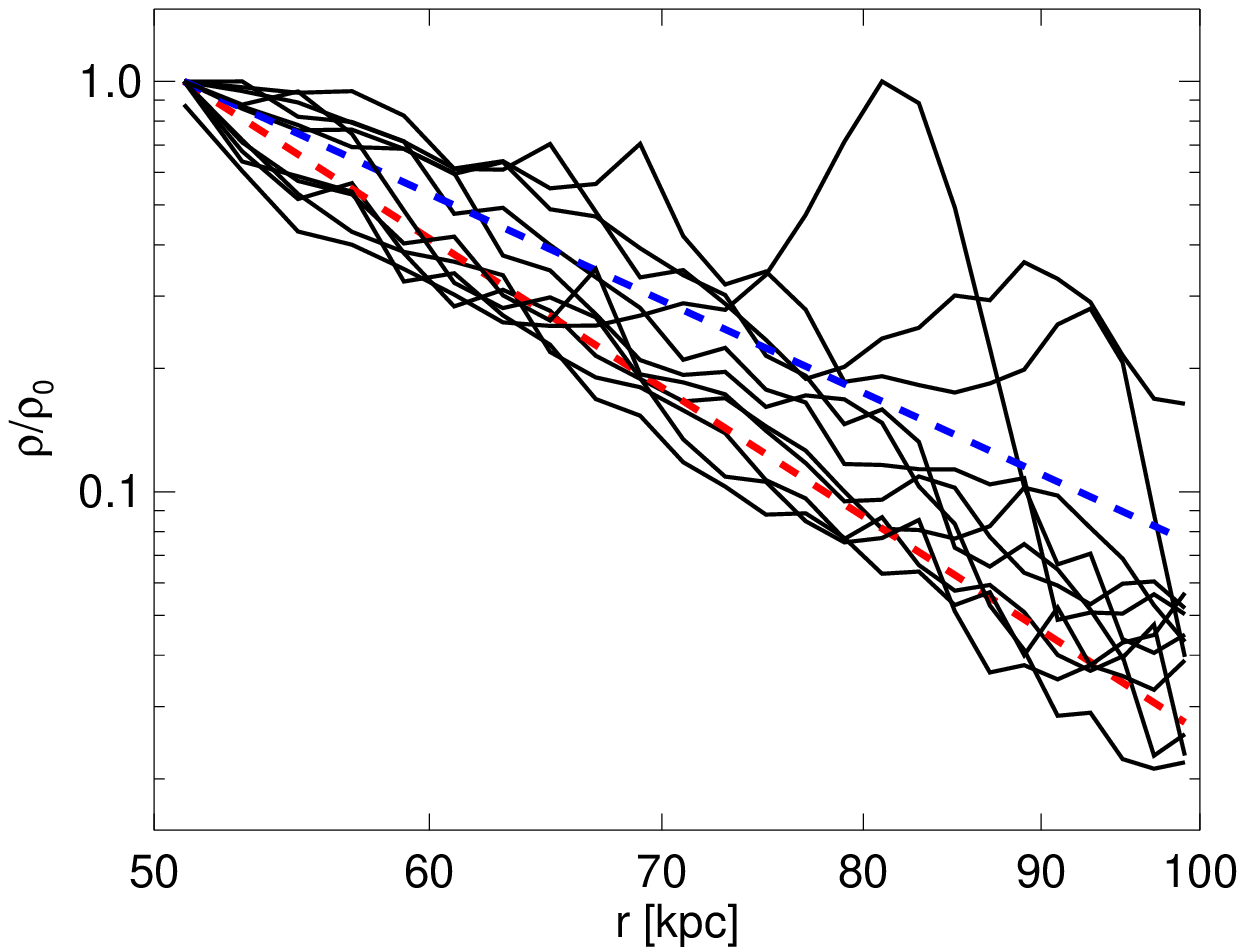}
   \end{minipage}
  \begin{minipage}{0.5\linewidth}
    \centering
   \includegraphics[width=8.5cm, height=6.8cm]{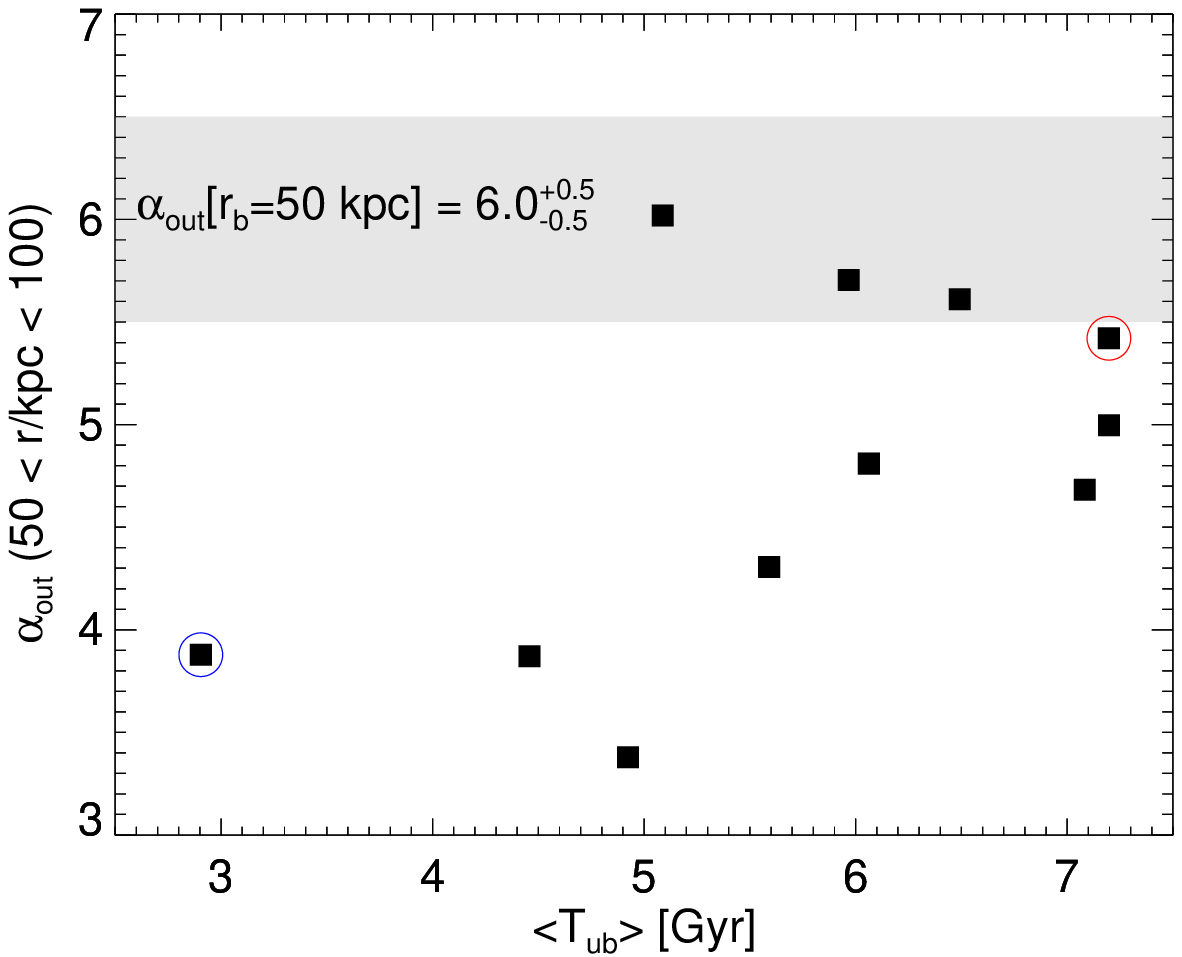}
   \end{minipage}
    \caption[]{\small \textit{Left panel:} Outer ($r > 50$ kpc) stellar halo density profiles of the 11 \cite{bullock05} simulations.  The dashed red and blue lines illustrate power-law fits of $\alpha \sim 5.4$ and $\alpha \sim 3.9$, respectively. \textit{Right panel:} The outer power-law slope as a function of average time that stars presently in the radial regime $50 < r/\mathrm{kpc} < 100$ became unbound from their parent dwarf ($T_{\rm ub}$) . The filled regions indicate the approximate slope for the Milky Way. Halos with shallower slopes tend to have more recent accretion activity.}
   \label{fig:bj}
\end{figure*}

\subsection{Milky Way accretion history}
Our finding of a strikingly sharp drop in stellar density beyond $r \sim 50-60$ kpc may have important implications for the accretion history of the stellar halo. In particular, our results suggest that, other than the relatively recent accretion of the SGR dwarf, the ``cannibalistic'' past of the Milky Way likely subsided several Gyr ago.

To illustrate the dependence of the outer stellar halo slope on its past accretion history, we compare with the \cite{bullock05} stellar halo models. This suite of 11 simulated stellar halos are built up entirely from the disruption of dwarf galaxies. The accretion history of each $M_{\rm vir} \sim 1.4 \times 10^{12}M_\odot$ mass halo is generated at random using semi-analytic merger trees appropriate for a $\Lambda$CDM cosmology. In the left-panel of Fig. \ref{fig:bj} we show the density profiles of the 11 halos between $50 < r/\mathrm{kpc} < 100$. It is clear that there is no ``universal'' outer halo fall-off, and there is a wide variation in the density profiles. In the right-panel of this Figure we show the outer stellar density slope against the average time at which the stars in the radial regime $50 < r/\mathrm{kpc} < 100$ became unbound from their parent dwarf ($T_{\rm ub}$). The filled gray region indicates the approximate slope for the Milky Way, $\alpha_{\rm out}[r_b = 50$ kpc$] =6$.

Despite some scatter, it is clear that halos with shallower slopes tend to have more recent accretion activity. It is worth noting that we have made no attempt to ``excise'' substructure from these simulated halos, and this will likely lead to steeper profiles in some cases. However, the general trend indicates that beyond $r \sim 50$ kpc in the Milky Way halo, the ``field'' halo stars were likely stripped from dwarfs that were accreted a long time ago ($> 6$ Gyr). 

Finally, we note that the Eris simulation (\citealt{guedes11}), one of the highest resolution hydrodynamical simulations of the formation of a $M = 8 \times 10^{11}M_\odot$ late-type spiral, also has a very steep fall-off in stellar density beyond $r \sim 60-70$ kpc (see \citealt{rashkov13} Fig. 2). This simulation, which has been successful in matching several Milky Way properties, has an early accretion history and high concentration ($c_{\rm vir} \sim 24$). This adds further weight to our deductions from the \cite{bullock05} simulations, that the Milky Way halo has undergone a relatively quiescent accretion history over the past several Gyr.

\subsection{Milky Way mass}
The total mass of the Milky Way halo remains a highly debated topic in the literature. In recent years, a surprising disparity has emerged between studies using the dynamics of halo stars to measure the total mass (e.g. \citealt{xue08}; \citealt{deason12}), and constraints based on satellite kinematics or timing arguments (e.g. \citealt{li08}; \citealt{boylan13}). The latter approaches tend to favor larger Milky Way masses ($> 1 \times 10^{12}M_\odot$) than the former ($< 1 \times 10^{12}M_\odot$).

However, the Jeans equations normally used to relate halo stars kinematics to total mass, suffer from strong degeneracies with the tracer velocity anisotropy and tracer \textit{density slope}. The mass-anisotropy degeneracy is well known, but the influence of the adopted tracer density profile is often ignored. \cite{dehnen06} stressed the importance of the tracer density profile by showing that a sharp drop in stellar density is able to reconcile relatively massive dark matter halo models with a declining velocity dispersion profile (see below). The line-of-sight (LOS) velocity dispersion profile of halo stars declines dramatically beyond $r \sim 50-60$ kpc (see \citealt{deason12} Fig.9). With LOS velocity information alone, it is not obvious whether this is due to a property of the tracers or the underlying mass profile. Our finding that the tracer density profile declines rapidly beyond $r \sim 50-60$ kpc suggests that this drop is caused, at least in part, by the tracer density profile.

\cite{deason12} show, under a range of assumptions about tracer properties, that the total Milky Way mass within 150 kpc lies between $5-10 \times 10^{11} M_\odot$. Steeper tracer density profiles push this constraint to the \textit{higher} mass end. We note that \cite{boylan13} state that, within the uncertainties, their constraint on the Milky Way mass using the 3D kinematics of Leo I, agree with \cite{deason12}, but only at the low mass end. Therefore, our finding of a rapidly declining stellar halo density profile, may play a large role in reconciling these apparently disparate Milky Way mass constraints.

However, it is premature to suggest that the issue is now resolved. The Eris simulation (mentioned above), has a low mass Milky Way halo ($\sim 8 \times 10^{11}M_\odot$), but its stellar halo also shows a rapid fall-off beyond $r \sim 60$ kpc. In the same radial regime, Eris has a steep dark matter mass profile and the halo stars have very radially biased orbits ($\beta \to 1$). Thus, the tracer density slope \textit{alone} cannot rule out a low-mass Milky Way halo. This emphasizes the importance of measuring the velocity anisotropy of distant halo stars. Thankfully, with the advent of the upcoming \textit{Gaia} mission and deep, multi-epoch \textit{HST} proper motion measurements (\citealt{deason13b}; \citealt{hstpromo}), this will be possible in the very near future.

\subsection{Comparison with M31}
\label{sec:m31}
Recent work by the SPLASH collaboration (Spectroscopic and Photometric Landscape of Andromeda's Stellar Halo; \citealt{gilbert12}) and the PAndAS team (The Pan-Andromeda Archaeological Survey; \citealt{ibata14}) have mapped the density profile of the M31 stellar halo out to remarkably large distances ($R \sim 200$ kpc). Both of these teams find that the stellar distribution can be described by a single power-law with $\alpha \sim 3-3.5$, and there is no evidence for a steepening at large radii. These results are in stark contrast to our findings for the Milky Way stellar halo, where the stellar density plunges dramatically beyond $r \sim 50$ kpc.

The shallower density slope of M31 halo stars suggests it has undergone a much more active late-time accretion history than the Milky Way (see above discussion and Fig. \ref{fig:bj}). This is in agreement with the conclusions of \cite{deason13a} who argue that the absence of a ``break'' (i.e. a transition to a steeper density profile at large radii) in the stellar density of the M31 halo suggests a more recent accretion history. 

\section{Conclusions}
We model the density distribution of distant BHB and BS halo stars using SDSS DR9 photometry, with the aim of measuring the outer slope of the Milky Way stellar halo density profile beyond $r \sim 50$ kpc.  We construct number density PDFs in $ugr, m_g$ space, and include contributions from QSO contaminants. Our PDF is convolved with the $ugr, m_g$ error distribution to take into account the significant photometric uncertainties at faint magnitudes. We fix the QSO number density using the QSO model developed by \cite{bovy11}, and allow the stellar halo profile within $r \sim 40-50$ kpc to lie within an observationally motivated parameter space. The outer halo model parameters are identified by modeling the stellar distribution in $u-g$, $g-r$, $m_g$, $\ell$, $b$ space. We test our method on simulated catalogs of BHBs, BSs and QSOs, and demonstrate that the properties of the distant halo can be recovered with sufficient accuracy.

We apply our likelihood analysis to high latitude ($|b| > 30$ deg) SDSS DR9 stars in the color and magnitude range; $0.7 < u-g < 1.6$, $-0.25 < g-r < -0.1$ and $18.5 < g < 20.5$. With this selection, BHB and BS stars span a heliocentric distance range: $10 \lesssim D_{\rm BS}/\mathrm{kpc} \lesssim 75$,  $40 \lesssim D_{\rm BHB}/\mathrm{kpc} \lesssim 100$. We identify stars coincident on the sky with the known substructures Virgo and SGR, and apply our analysis both including and excluding these stars. Our analysis assumes: 1) stellar halo sphericity at large radii, 2) an inner stellar halo ($r \lesssim 40$ kpc) density parametrization consistent with current constraints in the literature, and 3) BHB and BS intrinsic color distributions that remain the same throughout the halo.

The relative contributions of A-type stars (BHB and BS) and QSOs are computed iteratively from the convolved PDFs for each set of model parameters. In our selection box $0.7 < u-g < 1.6$, $-0.25 < g-r < -0.1$, and magnitude range $18.5 < g < 20.5$, we find a QSO contamination fraction of $f_{Q} \sim 0.2$ and a BHB fraction of $f_{\rm BHB} \sim 0.2$; these fractions have a weak dependence on our model parameters. 

After excluding known substructures, we find that very steep outer halo profiles are preferred, with $\alpha_{\rm out} \sim 6$ beyond $r=50$ kpc. Even when SGR and Virgo stars are included in the analysis, we find very steep outer profiles. There is evidence for a break in the stellar density at $r_b \sim 50-60$ kpc, which is coincident with the apocenter of the SGR leading arm.

We compare our results to the predictions of simulated stellar halos. The \cite{bullock05} suite of halos, built up purely from the accretion of dwarf galaxies, have outer profile slopes which depend on the accretion history of the halo; steeper outer slopes suggest earlier accretion epochs than shallow slopes. Thus, our finding of a very steep outer halo profile argues that, apart from the relatively recent accretion of SGR, the majority of the Milky Way stellar halo was built up from relatively early accretion events ($ T > 6$ Gyr ago). This is in contrast to the M31 stellar halo which has a much shallower density slope out to $r \sim 200$ kpc ($\alpha \sim 3-3.5$; \citealt{gilbert12}; \citealt{ibata14}), and thus presumably has had a more active late-time accretion history.

The density profile of the Milky Way stellar halo is an important ingredient for dynamical mass estimates of the Galaxy. Until now, the unknown stellar density slope beyond $\sim 50$ kpc has proved to be a troublesome bottleneck in constraining the total mass out to large distances. Our finding of a very steep outer halo slope may have important implications for studies utilizing the kinematics of halo stars to estimate the total mass of the Milky Way.
The measurement we report here, in combination with constraints on the halo star velocity anisotropy from upcoming surveys (such as \textit{Gaia}) will, undoubtedly, significantly reduce the uncertainty surrounding dynamical mass measurements of the Milky Way.

\section*{Acknowledgments}
We thank Jay Farihi for his help with the white dwarf models. AJD is currently supported by NASA through Hubble Fellowship grant HST-HF-51302.01, awarded by the Space Telescope Science Institute, which is operated by the Association of Universities for Research in Astronomy, Inc., for NASA, under contract NAS5-26555. VB thanks the Royal Society and the European Research Council for financial support.  We thank the Aspen Center for Physics and the NSF Grant \#1066293 for hospitality during the conception of this paper. We also thank an anonymous referee for useful comments.

\begin{appendix}
\section{White dwarf models}
\label{sec:wd}

\begin{figure*}
  \begin{minipage}{0.66\linewidth}
    \centering
    \includegraphics[width=12cm, height=4cm]{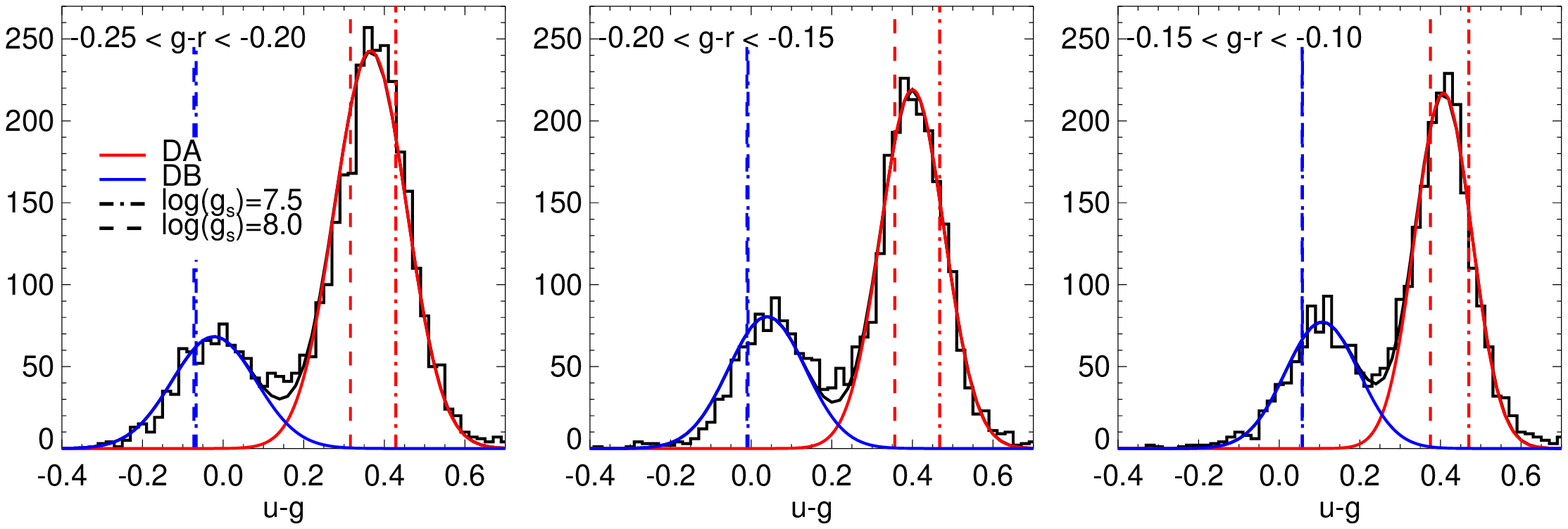}
   \end{minipage}\hspace{1pt}
  \begin{minipage}{0.33\linewidth}
    \centering
   \includegraphics[width=6cm, height=3.43cm]{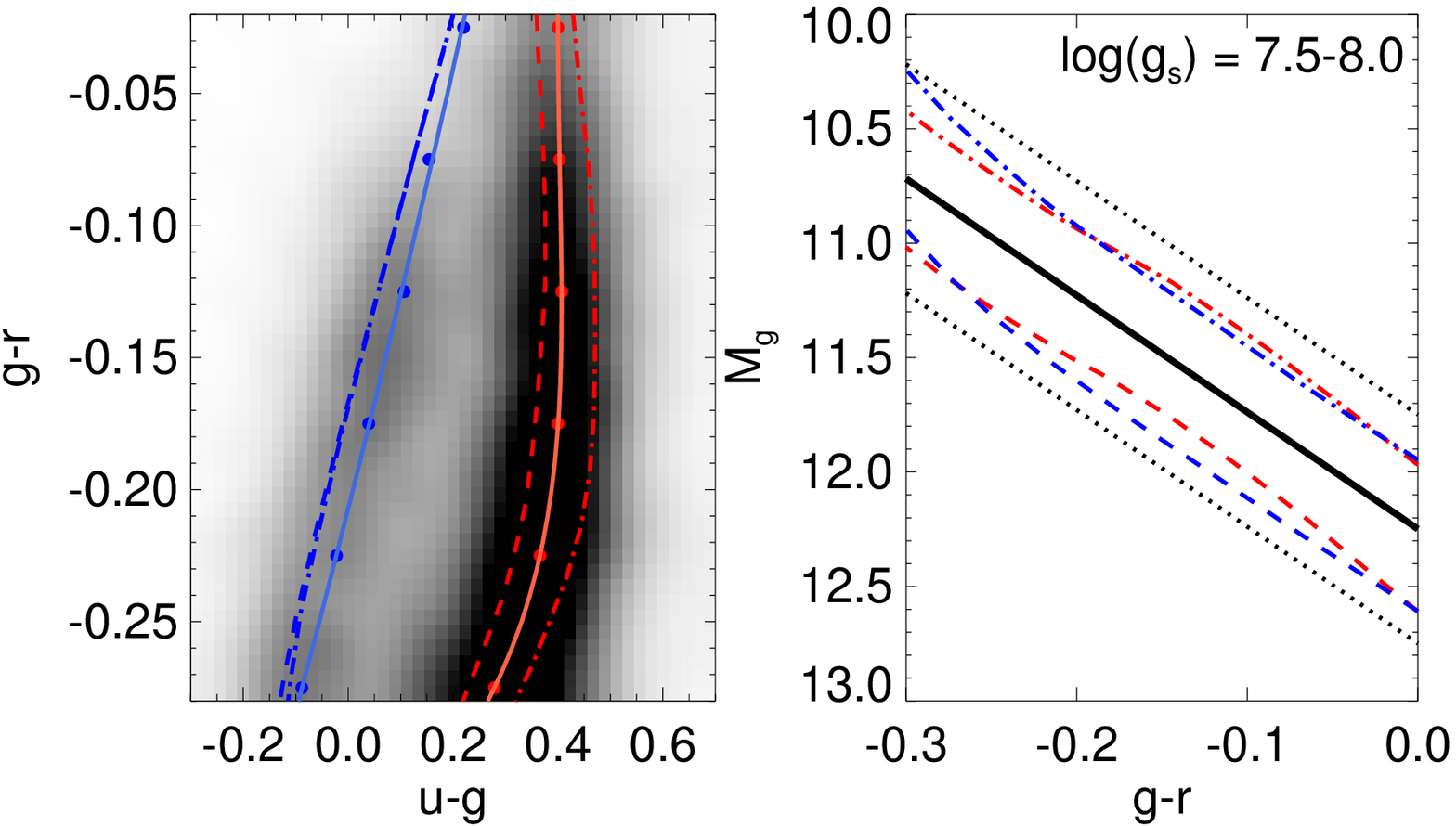}
   \end{minipage}
    \caption[]{\small \textit{Left three panels:} The distributions of white dwarf (WD) stars in $u-g$ color space in 3 different $g-r$ color bins. The two populations are DA (red, DA = H-dominated atmosphere) and DB (blue, DB = He-dominated atmosphere) type WDs. The dashed (dot-dashed) lines indicate the peak positions for model WDs with surface gravity $\mathrm{log}(g_s)=8.0(7.5)$. \textit{Right two panels:}  The first panel shows WDs in $u-g, g-r$ color space in the magnitude range $18.5 < g < 20.5$. The filled circles indicate the ``peak'' $u-g$ values for the DA and DB populations shown in the left panels. The solid lines show polynomial fits, defining the WD $u-g, g-r$ ridgelines. The dashed (dot-dashed) lines indicate the WD ridgelines for models with $\mathrm{log}(g_s) =8.0(7.5)$. The second panel shows the relation between WD absolute magnitudes and $g-r$ color in the $\mathrm{log}(g_s) =8.0(7.5)$ models. The red and blue lines are for DA and DB type WDs respectively. These DA- and DB-type sequences have very similar absolute magnitudes. We use the weighted mean of these sequences (indicated by the black line) as our absolute magnitude calibration. The dotted lines indicate a spread of 0.5 mag about this relation.}
   \label{fig:wd}
\end{figure*}

We select white dwarfs (WDs) by applying the XDQSO algorithm (see \S\ref{sec:qso_models}) to our high latitude ($|b| > 30^\circ$) SDSS DR9 photometry in the magnitude range $18.5 < g < 20.5$, and only consider blue stars ($u-g < 0.7$) with a low QSO probability ($p_{\rm qso} < 0.1$).

The left three panels of Fig. \ref{fig:wd} show the distribution of WD stars in $u-g$ color space for three $g-r$ color bins. Two populations are evident: these are the H-dominated (redder $u-g$) and He-dominated (bluer $u-g$) populations, which we refer to as DA and DB type respectively. The solid lines show a double Gaussian fit to these populations. The vertical dashed (dot-dashed) lines indicate the loci of model white dwarfs with surface gravity $\mathrm{log}(g_s)=8.0(7.5)$. These model white dwarfs derive from the Montreal WD atmosphere group\footnote{\url{http://www.astro.umontreal.ca/$\sim$bergeron/CoolingModels}} (\citealt{holberg06}; \citealt{kowalski06}; \citealt{tremblay11}; \citealt{bergeron11}), who provide synthetic model sequences for WDs with SDSS $ugriz$ photometry.

In the two right panels of Fig. \ref{fig:wd} we show the ridgelines of the WD populations in $ugr$ space based on the double Gaussian fits. The dashed lines indicate the ridgelines predicted by models with $\mathrm{log}(g_s)=7.5-8.0$. The agreement is good, especially for the DA-type dwarfs which dominate the WD population. We define the WD ridgelines with polynomials (cf. eqn \ref{eq:ridge}):
\begin{eqnarray}
(u-g)^{0}_{\rm DA}&=& 0.400+0.046(g-r)+1.186(g-r)^2 -2.227(g-r)^3-42.785(g-r)^4 \notag \\
(u-g)^0_{\rm DB}&=& 0.245+1.098(g-r)-0.412(g-r)^2 
\end{eqnarray}
for $-0.25 < g-r < -0.1$. 

We calculate the intrinsic spread of the two populations about their ridgelines to be $\sigma_{\rm  DA,0}(u-g)=0.060$ and $\sigma_{\rm  DB,0}(u-g)=0.075$. In a similar fashion to the A-type stars, we assume Gaussian distributions about the ridgelines:

\begin{eqnarray}
\label{eq:prob_wd}
p(u-g|{\rm DA, g-r})\propto\mathrm{exp}\left(-\frac{\left[(u-g)-(u-g)_{\rm DA}^0\right]^2}{2\sigma_{\rm DA}^2}\right), \notag\\ 
p(u-g|{\rm DB, g-r})\propto\mathrm{exp}\left(-\frac{\left[(u-g)-(u-g)_{\rm DB}^0\right]^2}{2\sigma_{\rm DB}^2}\right).
\end{eqnarray}

We assume a constant intrinsic $g-r$ distribution, so the color-based probabilities of class membership are then: $p(ugr| \mathrm{DA}) \propto p(u-g|{\mathrm{DA}, g-r})$, $p(ugr| \mathrm{DB}) \propto p(u-g|{\mathrm{DB}, g-r})$. We fix the fraction of DA-type white dwarfs (assuming just DA and DB types) to be $f_{DA} =0.7$.

In the right-most panel of Fig. \ref{fig:wd} we show the absolute magnitude-color relation for the model white dwarfs (with $\mathrm{log}(g_s)=7.5-8.0$). The DA and DB-types have similar absolute magnitudes. We adopt the weighted mean of these relations ($f_{DA} =0.7, f_{DB}=0.3$) as the average WD absolute magnitude relation, and assume a 0.5 mag spread to account for uncertainties in log($g_s$) (e.g.  log($g_s) = 7.5(8.0)$ models have brighter(fainter) absolute magnitudes by $\sim 0.5$ dex):
\begin{equation}
M_{g(\rm WD)}= 12.249 + 5.101(g-r)
\end{equation}
where, $\sigma_{M_g(\rm WD)} \sim 0.5$. 

Finally, we fix the WD density distribution assuming a disk distribution of stars. We use the disk density profile found by \cite{juric08}, which assumes an exponential profile and has contributions from thin and thick disk populations:
{\setlength\arraycolsep{0.1em}
\begin{eqnarray}
\rho_{\rm thin} &=& \mathrm{exp}\left(R_0/L1\right) \mathrm{exp}\left(-R_{\rm WD}/L1-|z_{\rm WD}+z_0|/H1\right) \notag \\
\rho_{\rm thick} &=& \mathrm{exp}\left(R_0/L2\right) \mathrm{exp}\left(-R_{\rm WD}/L2-|z_{\rm WD}+z_0|/H2\right) \notag \\
\rho_{\rm WD} &=& \rho_{\rm thin}+0.12\rho_{\rm thick}
\end{eqnarray}
}
where, $H1=0.3, L1=2.6, H2=0.9, L2=3.6, z_0=0.025$ kpc, $R_0=8.5$ kpc.

Fig. \ref{fig:ug_contam} in the main text shows that our WD models predict a very small fraction of WDs in our A-type star selection box, so we do not consider their contribution in our modeling procedure.

\section{Flattening and inner stellar halo density profile}
\label{sec:ftest}

\begin{figure*}
  \begin{minipage}{0.6\linewidth}
    \centering
    \includegraphics[width=6.5cm, height=12cm]{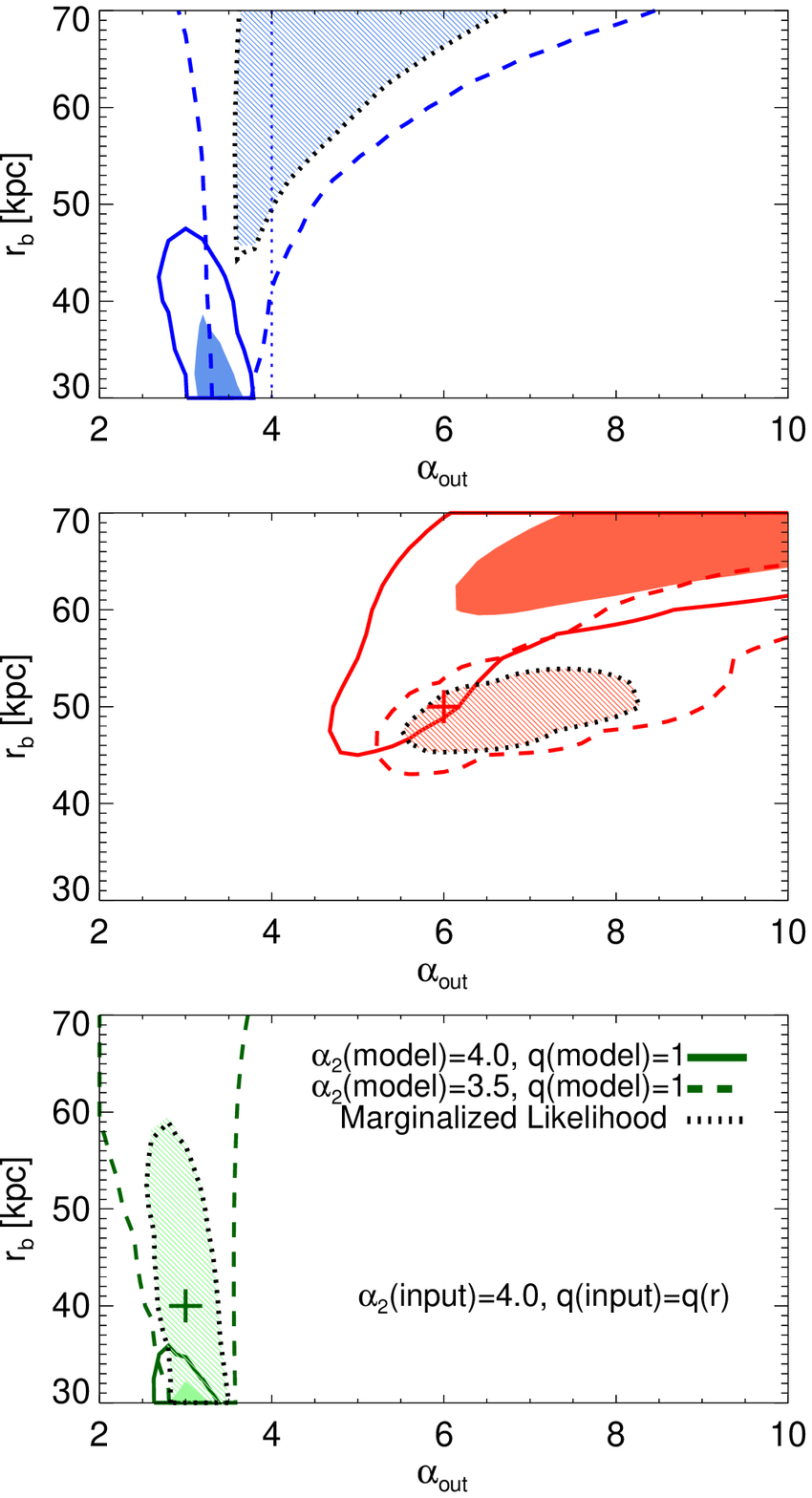}
   \end{minipage}\hspace{-100pt}
  \begin{minipage}{0.6\linewidth}
    \centering
   \includegraphics[width=6.5cm, height=12cm]{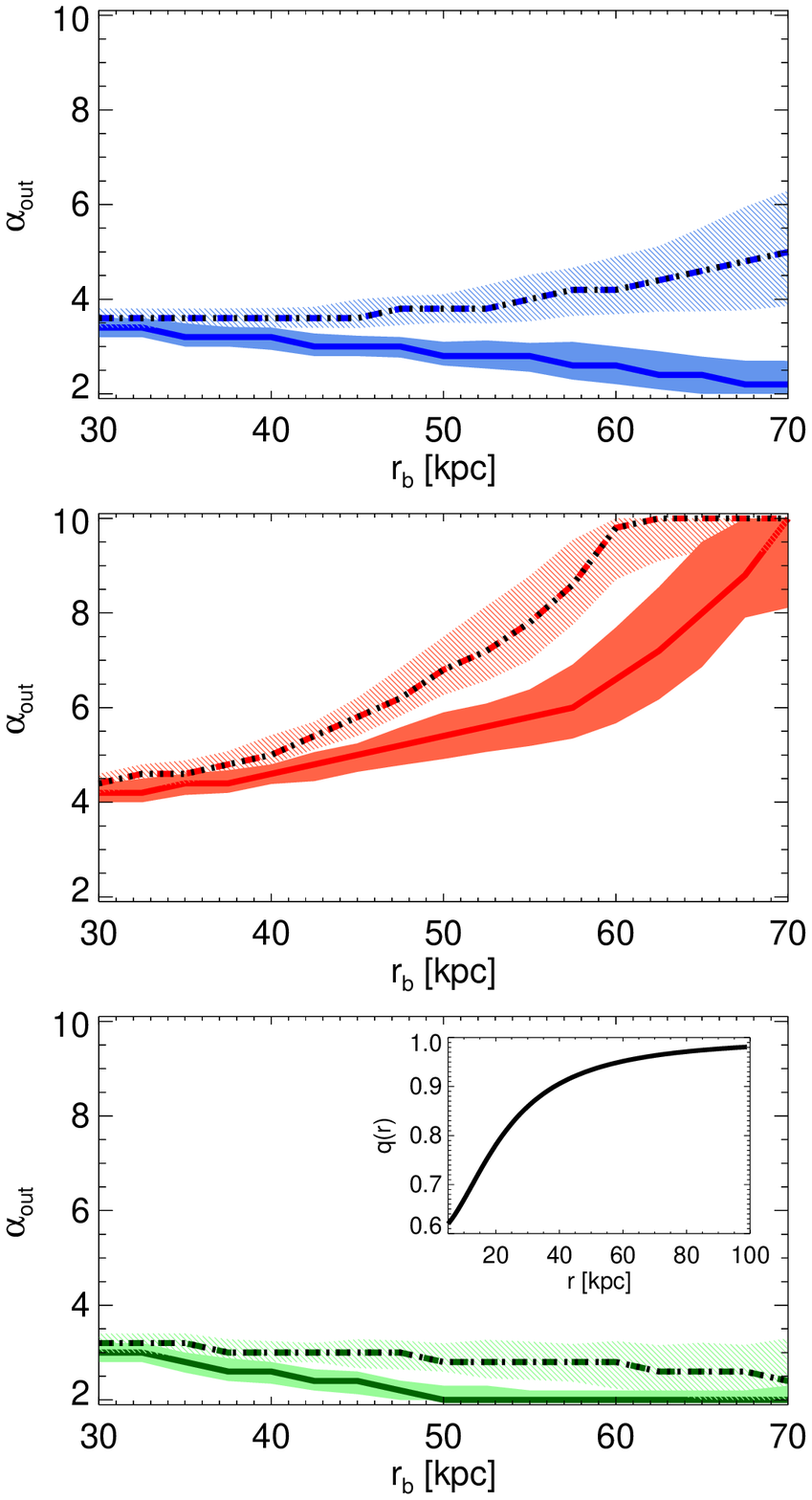}
   \end{minipage}
    \caption[]{\small Likelihood contours. The blue, red and green contours indicate the constant ($\alpha_{\rm out}=4.0, r_b=r_c$), steep ($\alpha_{\rm out}=6.0, r_b=50$ kpc) and shallow ($\alpha_{\rm out}=3.0, r_b=40$ kpc) toy models, respectively. In all cases, mock data are generated for an inner stellar halo profile with $\alpha_1=2.5, \alpha_2=4.0, r_c=25$ kpc (see eqn. \ref{eq:model}), and a minor-to-major axis ratio, $q$, that varies with radius ($q=q(r)$, see main text). Solid filled regions and solid lines indicate the 1- and 2-$\sigma$ confidence regions when a spherical model ($q=1$) is used in the likelihood analysis with inner stellar halo parameters: $\alpha_1=2.5, \alpha_2=4.0, r_c=25$ kpc. Similarly, line-filled regions and dashed lines indicate the 1- and 2-$\sigma$ confidence regions for a spherical model with inner stellar halo parameters: $\alpha_1=2.5, \alpha_2=3.5, r_c=25$ kpc. The dotted black lines indicate the 1-$\sigma$ contour after marginalizing over the two inner density profile models; the marginalized likelihood is dominated by the higher likelihood model. \textit{Right panels:} The maximum likelihood outer slope for different fixed values of $r_b$. The lines show the maximum likelihood parameters, and the shaded regions indicate the 1-$\sigma$ uncertainties. The line-styles and colors are the same as the left-panels. The inset in the bottom-right panel shows the radial dependence or the flattening parameter, $q(r)$ that we adopt for this exercise.}
   \label{fig:ftest}
\end{figure*}

Our analysis assumes spherical stellar halo density profiles. Here, we consider the implications of this assumption for non-spherical profiles. Mock data is generated, as described in \S\ref{sec:mock}, but our model stellar halo profiles are given a minor-to-major axis ratio, $q$, which varies with radius. We adopt the following parametrization for $q$:
\begin{equation}
q(r)=q_0 \sqrt{\frac{r^2+r_s^2}{\left(q_0 r\right)^2 +r_s^2}}
\end{equation}
and set $q_0=0.6$ and $r_s=15$ kpc. The minor-to-major axis parameter thus varies smoothly from $q \sim 0.6$ at small radii to $q \sim 1.0$ at large radii (see inset in bottom-right panel of Fig. \ref{fig:ftest}). Three mock datasets are generated, with the same parameters as described in \S\ref{sec:mock}. We apply our likelihood analysis to this simulated data \textit{assuming sphericity at all radii}.

The results of this exercise are summarized in Fig. \ref{fig:ftest} (cf. Fig \ref{fig:like_fake}). The ``constant'' (blue), ``steep'' (red) and ``shallow'' (green) toy models are shown in the top, middle and bottom rows respectively. Solid filled regions and solid lines indicate the 1- and 2-$\sigma$ confidence regions when a spherical model ($q=1$) is used in the likelihood analysis with inner stellar halo parameters: $\alpha_1=2.5, \alpha_2=4.0, r_c=25$ kpc. In all cases, the resulting outer stellar density parameters ($r_b$ and $\alpha_{\rm out}$) are biased towards shallower profiles: typically $\alpha_{\rm out}$ is 0.5 dex too shallow. Our flattened model mainly affects the BS stars at small radii; when forced to fit to a spherical model, the BS distribution appears shallower than the input inner density profile ($\alpha_2(\mathrm{input})=4.0$). The best-fit model, with only $\alpha_{\rm out}$ and $r_b$ as free parameters, compensates for this bias by making $\alpha_{\rm out}$ slightly shallower.

The line-filled regions and dashed lines indicate the 1- and 2-$\sigma$ confidence regions when a spherical model ($q=1$) is used in the likelihood analysis with inner stellar halo parameters: $\alpha_1=2.5, \alpha_2=3.5, r_c=25$ kpc. Thus, this model adopts a shallower inner profile ($\alpha_2=3.5$) than the case described above. In this case, we are able to reproduce the correct outer stellar halo parameters with reasonable accuracy. By using a shallower inner profile in the modeling, we have ``compensated'' for the affect of flattening. So, we are able to reproduce the correct stellar density profile at large radii, even though we have neglected the affects of flattening. 

When we apply our analysis to the SDSS DR9 data, we marginalize our likelihood distribution over a wide range of inner profile parameters (see eqn. \ref{eq:inner_range}). This ensures that our inner profile is flexible enough to compensate for affects such as a flattened inner profile. The dotted black lines in the left-panels of Fig. \ref{fig:ftest} indicate the 1-$\sigma$ contours after marginalizing over the two inner density profile models. It is clear that the marginalized likelihood is dominated by the (higher likelihood) model which is able to reproduce the correct outer stellar density slope. This gives us confidence that the resulting outer stellar density parameters are robust to variations in the inner stellar halo profile. 

Note that we also use mock data to test how variations in $\alpha_1$ may affect our results (this is kept fixed at $\alpha_1=2.5$ in our analysis) . We find that the outer profile parameters are less sensitive to variations in $\alpha_1$ than $\alpha_2$ or $r_c$, and we find little difference if $\alpha_1$ is changed by $\sim \pm 0.5$ dex.

\end{appendix}

\end{document}